\documentclass{AIAA}
\usepackage{amssymb}
\usepackage{amsmath}
\usepackage{longtable}
\usepackage{color}
\usepackage[normalem]{ulem} 

\begin{document}

\title{Transition Prediction for Scramjet Intakes Using the $\gamma$-$Re_{\theta_t}$ Model Coupled to Two Turbulence Models}
\def\baselinestretch{2}
\author{Sarah Frauholz\footnote{Research Fellow, Chair for Computational Analysis of Technical Systems, Center for Computational Engineering Science (CCES), RWTH Aachen, Schinkelstr. 2, 52062 Aachen, frauholz@cats.rwth-aachen.de.}} 
\author{Birgit U.~Reinartz\footnote{Research Associate, Chair for Computational Analysis of Technical Systems (CATS), CCES, Senior Member AIAA, RWTH Aachen, Schinkelstr. 2, 52062 Aachen, reinartz@cats.rwth-aachen.de.}}
\author{Siegfried M\"uller\footnote{Professor, Institut f\"ur Geometrie und Praktische Mathematik, RWTH Aachen, Templergraben 55, 52062 Aachen, mueller@igpm.rwth-aachen.de.}}
\author{Marek Behr\footnote{Professor, Chair for Computational Analysis of Technical Systems (CATS), CCES, RWTH Aachen, Schinkelstr. 2, 52062 Aachen, behr@cats.rwth-aachen.de.}}

\affiliation{\vspace*{15pt} \\RWTH Aachen University, Germany} 
\def\baselinestretch{1}
\begin{abstract}
Due to the thick boundary layers in hypersonic flows, the state of the boundary layer significantly influences the whole flow field as well as surface heat loads. Hence, for engineering applications the efficient numerical prediction of laminar-to-turbulent transition is a challenging and important task. 
Within the framework of the Reynolds averaged Navier-Stokes equations, Langtry/Menter~\cite{menter_transition3}  proposed the $\gamma$-$Re_{\theta_t}$ transition model using two transport equations for the intermittency and $Re_{\theta_t}$ combined with the Shear Stress Transport turbulence model (SST)~\cite{menter:94}. The transition model contains two empirical correlations for onset and length of transition. Langtry/Menter~\cite{menter_transition3} designed and validated the correlations for the subsonic and transonic flow regime. For our applications in the hypersonic flow regime, the development of a new set of correlations proved necessary, even when using the same SST turbulence model~\cite{martin}.
Within this paper, we propose a next step and couple the transition model with the SSG/LRR-$\omega$ Reynolds stress turbulence model~\cite{Eisfeld:06} which we found to be well suited for scramjet intake simulations. 
First, we illustrate the necessary modifications of the Reynolds stress model and the hypersonic in-house correlations using a hypersonic flat plate test case. Next, the transition model is successfully validated for its use coupled to both turbulence models using a hypersonic double ramp test case. Regardless of the turbulence model, the transition model is able to correctly predict the transition process compared to experimental data. In addition, we apply the transition model combined with both turbulence models to three different fully 3D scramjet intake configurations which are experimentally investigated in wind tunnel facilities. The agreement with the available experimental data is also shown.
\end{abstract}

\maketitle

\section*{Nomenclature}
\noindent\begin{longtable}{@{}lcl@{}}
$c_{p}$ & : & Specific heat at constant pressure, pressure coefficient [-]\\
$\delta_{ij}$ & : & Kronecker Delta [-]\\
$E$ & : & Specific total energy [m$^2$/s$^2$]\\
$\varepsilon_{thres}$& : & Threshold value used for data compression [-]\\
$\varepsilon_l$& : & Level-dependent threshold value for level $l$ [-]\\
$H$ & : & Total specific enthalpy [m$^2$/s$^2$]\\
$I$& : & Turbulent intensity [-]\\
$k$ & : & Turbulent kinetic energy [m$^2$/s$^2$]\\
$L$ & : & Maximum refinement level [-]\\
$l$ & : & Local refinement level [-]\\
$\mu$ & : & Molecular viscosity [kg/(m s)]\\
$\mu_l$ & : & Laminar viscosity [kg/(m s)]\\
$\mu_t$ & : & Turbulent viscosity [kg/(m s)]\\
$\omega$ & : & Specific turbulence dissipation rate [1/s]\\
$p$ & : & Pressure [Pa]\\
$p_t$ & : & Total pressure [Pa]\\
$q_{i}$ & : & Component of heat flux vector [W/m$^2$] \\
$q_k^{(t)}$ & : & Turbulent heat flux [W/m$^2$] \\
$\rho$ & : & Density [kg/m$^3$]\\
$St$& : & Stanton number [-]\\
$t$ & : & Time [s]\\
$T$& : & Temperature [K]\\
$T_{\rm w}$ & : & Wall temperature [K] \\
$T_{0}$ & : & Total temperature [K] \\
$u_i$& : & Velocity component [m/s]\\
$x_i$& : & Cartesian coordinates component [m]\\
$x$, $y$, $z$ & : & Cartesian coordinates [m]\\
$y^+$ & : & Dimensionless wall distance [-]\\
$M$ & : & Mach number [-]\\
$Re$ & : & Reynolds number [1/m]\\
$Res_{drop}$& : & Averaged density residual, at which the adaptations are performed [-]\\
$D_{ij}$ & : & Diffusion tensor for the Reynolds stresses [m$^2$/s$^3$]\\
$\epsilon_{ij}$& : & Destruction tensor for the Reynolds stresses [m$^2$/s$^3$]\\
$M_{ij}$ & : & Turbulent mass flux tensor for the Reynolds stresses [m$^2$/s$^3$]\\
$\Pi_{ij}$ & :  & Re-distribution tensor for the Reynolds stresses [m$^2$/s$^3$]\\
$P_{ij}$ & : & Production tensor for the Reynolds stresses [m$^2$/s$^3$]\\
$\tilde{R}_{ij}$ & : & Reynolds stress tensor [m$^2$/s$^2$]\\
$\tau_{ij}$ & : & Viscous stress tensor [m$^2$/s$^2$]\\
%
$\gamma$&:& Intermittency [-]\\
$\gamma_{eff}$&:& Effective intermittency [-]\\
$ Re_{\theta t}$&:& transition onset Reynolds number [-]\\
$Re_{\theta_c}$&:& Critical Reynolds number, empirical correlation [-]\\
$F_{length}$&:& transition length function, empirical correlation [-]\\
$P_\gamma$&:& Production term of the $\gamma$ transport equation [m/s]\\
$E_\gamma$&:& Destruction term of the $\gamma$ transport equation [m/s]\\
$P_{\theta_t}$&:& Production term of the $Re_{\theta t}$ transport equation [m/s]\\
$\frac{\partial \cdot}{\partial \cdot}$ & : & Partial derivative \\
$\overline{\cdot}$ & : & Reynolds-averaged quantity \\
$\tilde{\cdot}$ & : & Favre-averaged quantity \\
$\cdot_\infty$ & : & Freestream value 
\end{longtable}
\section{Introduction}
\label{sec:intro}
The study of hypersonic flows has been of interest for more than six decades~\cite{martin}. Nowadays, a major application in the field of hypersonics is the realization of a supersonic combustion ramjet (scramjet), an airbreathing propulsion system that operates above Mach 5 and at approximately 30-40 km altitude. One major impediment to the realization of such an engine lies in the uncertainties related to its aerothermodynamic design. The study of hypersonic configurations at real flight conditions is both experimentally as well as numerically demanding, though not for the same reasons. On the one hand, hypersonic test facilities need a huge amount of energy to establish high-enthalpy flow conditions. Short duration test times and vitiated air effects are just two of the resulting drawbacks. On the other hand, numerical simulations have to deal with modelling uncertainties with respect to turbulence, transition and high temperature effects as well as limited computer resources. 

Up to now, turbulent flow simulations for hypersonic engineering applications at realistic Reynolds numbers are only computationally affordable when applying the Reynolds Averaged Navier-Stokes (RANS) equations. The most widely used turbulence models in this field are the eddy viscosity models, where a linear dependence between the Reynolds stress tensor and the strain rate tensor is assumed. However, several literature reviews showed that these models perform poorly for wall dominated flows characterized by a thick boundary layer, a strong shock-wave-boundary-layer interaction and separation \cite{roy-blottner}, which are all typical for hypersonic applications. To overcome this deficiency, differential Reynolds stress turbulence models (RSM) can be applied. This class of models has not been widely used because of its decreased stability and the increased computational cost due to the presence of seven equations that describe turbulence. However, in an earlier study, the RSM was successfully used for the simulation of separated hypersonic boundary layer flow where common two-equations eddy viscosity models failed~\cite{Bosco:11b, Bosco:11}. 

\begin{figure*}[!ht]
\begin{center}
   \includegraphics[width=1\textwidth,clip ]{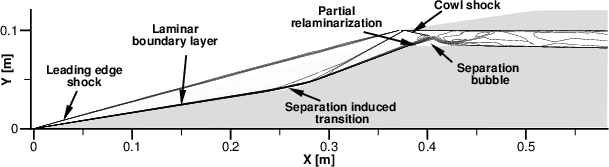}
   \caption{Two-dimensional adaptive computation of a scramjet intake (see \ref{Sec:SWL}) showing the main physical flow phenomena via Mach number lines.}
   \label{fig:2d-swl}	
\end{center}
\end{figure*}

For a scramjet intake, the state of the boundary layer (laminar, transitional, turbulent) plays an important role affecting, e.g., the size and location of  flow separation, the surface heat loads and the intake performance in terms of captured mass flow. 
%
For intakes with several ramps (see Fig. \ref{fig:2d-swl}) separation-induced transition is most likely to occur. Here, the laminar boundary layer separates at the end of the first ramp and within the shear layer over the separation bubble the flow transitions due to inviscid instability mechanisms~\cite{Lodefier:05}. The size of the separation bubble is reduced by the transition process within the shear layer~\cite{Krause:08b}.

For single ramp intakes, natural transition happens at the external ramp or side walls. Disturbances from the freestream and the wall 
are influencing the laminar boundary layer. At a critical Reynolds number, these disturbances are not damped any more and transition occurs~\cite{Mayle:91}. This transition process involves the generation of Tollmien-Schlichting waves followed by three-dimensional waves and vortex structures leading to vortex break-up and turbulent flow~\cite{Schlichting}. 
The Tollmien-Schlichting waves are first mode instabilities~\cite{McKeel_PhD_thesis} that occur at freestream turbulent intensities of less than 1\%, which are typical for hypersonic test facilities. 

Due to expansion corners, reverse transition (partial relaminarization) can occur as well~\cite{Krause:08b}. When the boundary layer experiences a positive pressure gradient in cross-flow direction and a negative pressure gradient in flow direction caused by a convex corner, the turbulent vortex movement is damped~\cite{Arnette:95}. Caused by the expansion of the boundary layer, the heat transfer is reduced slowing the turbulent mixing process and the flow partially relaminarizes~\cite{Maestrello:89}.

Due to the different physical mechanisms triggering transition, the development of transition models is a challenging task.
A common engineering approach to model transition is the use of fixed transition points~\cite{Rein:07b,Nguyen:2013}, which is computationally cheap and easy to implement. However, often the transition point is not known a priori, e.g., from experiments~\cite{Reinartz:12}. Hence, advanced methods that resolve the transition process are necessary. 

The semi-empirical $e^n$-method is one of the state of the art models for predicting natural transition. It is extensively used for industrial aircraft applications~\cite{Stock:00}. The $e^n$-method is based on local linear stability theory assuming parallel flow. The growth of the disturbance amplitude is computed from the boundary layer neutral point to the transition location~\cite{phd_thesis_langtry}. Therefore the mean flow is calculated at a large number of streamwise locations along the surface and at each location the spatial amplification rate for each unstable frequency is analysed using a local linear stability analysis or the parapolised stability equations. Then, the spatial amplification rate is integrated in streamwise direction on the body resulting in the amplitude ratio for each frequency. From this, the $n$-factor is defined as the maximum of the amplitude ratio at each streamwise location. When the disturbance ratio $e^n$ exceeds the limiting $n$-factor, the flow transitions
~\cite{warren:98a}. The main drawback of this method is that standard CFD solvers are not accurate enough and therefore the use of an additional boundary layer code is necessary. The assumptions of the linear stability theory and the difficulties to predict the growth of the disturbance amplitude ratio for three-dimensional flows~\cite{Stock:06} are additional limitations of this method.

Another approach is the use of empirical correlations for the critical Reynolds number such as Mayle~\cite{Mayle:91} and Abu-Ghannam~\cite{Abu_Ghannam:80}. This method computes a laminar solution as a first step. Then the boundary layer edge is detected using, e.g., the algorithm of Lodefier~\cite{Lodefier:05}. This is the most challenging part. Next, the momentum thickness $\theta$ and the momentum thickness Reynolds number $Re_\theta$ at the boundary layer edge are determined. If the momentum thickness Reynolds number $Re_\theta$ exceeds the critical Reynolds number computed by the empirical correlations, the turbulence model is applied. Examples for this type of models can be found in~\cite{Lodefier:06,Pecnik:03}. Some models consider the transition length as well~\cite{Mayle:91,Simeonides:06}. The main drawback of this method is the non-locality due to the integration of the values at the boundary layer edge. Therefore, its integration into a parallel CFD code is not straight forward.

Most transition models, such as the model of Steelant~
\cite{steelant:01}, the model of Suzen
~\cite{suzen:03}, the model of Papp
~\cite{papp:08} and the model of Warren
~\cite{warren:98a}, are based on non-local variables as well. Other models are not validated for the transition processes occuring within hypersonic intake flows, such as the model of Walters/Leylek~\cite{Walters:04,Walters:05} that is not validated for separation-induced transition.
%


Therefore, we chose the $\gamma$-$Re_{\theta_t}$ model by Langtry/Menter~\cite{menter_transition1,menter_transition2,menter_transition3}. 
It describes the transition process using the intermittency parameter $\gamma$. This parameter gives information about the fraction of time when the flow becomes turbulent and was first introduced by Dhawan/Narasimha during investigations of the transition on flat plates~\cite{dhawan:58}. 
In contrast to many other transition models, the $\gamma$-$Re_{\theta_t}$ model is based only on local variables by using the vorticity Reynolds number instead of the momentum thickness Reynolds number. Thus, the model can easily be integrated within a parallel CFD code. In addition, the model has a modification to account for separated flow transition.
The $\gamma$-$Re_\theta$ model is based on two transport equations for the intermittency and the transition onset criteria using the vorticity Reynolds number. Within the model, two empirical correlations for the transition onset Reynolds number $Re_{\theta_c}$ and the transition length function $F_{length}$ allow to calibrate the model towards different flow regimes.

Malan~\cite{malan:09} combined the $\gamma$-$Re_{\theta_t}$ transition model with the $k$-$\omega$ model of Wilcox~\cite{wilcox:08} and calibrated the empirical correlations for transonic and subsonic flows within the TRACE code. Grabe/Krumbein~\cite{grabe:13} implemented the transition model combined with the SST model within the DLR TAU code and introduced a modification for cross-flow transition. Watanabe also modified the transition model for crossflow transition~\cite{watanabe:09}. In addition, You~\cite{you:12} proposed a new equation for the effective intermittency for hypersonic flows taking the pressure gradient via the acceleration parameter $\lambda_\theta$ into account. Medida/Baeder combined the transition model with the Spalart Allamares turbulence model~\cite{Medida:11}. 
In~\cite{menter:14}, Menter proposed to improve the $\gamma$-$Re_{\theta_t}$ transition model by (1) reducing it to a single transport equation for the intermittency; (2) simplifing the empirical correlation to allow an easier fine-tuning; (3) a Galilean invariant formulation; and (4) including a modification to predict \textbf{cross-flow }instabilities.

Krause implemented the transition model into our flow solver and developed correlations for hypersonic flows~\cite{martin}.
Now, within this paper we exchange the SST turbulence model with the RSM turbulence model which improves the numerical prediction of separated hypersonic boundary layer flow~\cite{Bosco:11b, Bosco:11}. The hypersonic, in-house correlations within the transition model have to be calibrated for the RSM model.

The physical modelling with a special emphasis on the chosen transition model and its coupling with both turbulence models is described in Section \ref{sec:pm}. Subsequently, the numerical methods employed for the solution of the discrete problems are illustrated in Section~\ref{sec:qf}. In Section~\ref{sec:mod-trans-rsm}, the modifications of the in-house correlations for the transition model coupled to the RSM model (RSM-Tr) using a hypersonic flat plate test case is shown. Within the next Section~\ref{sec:validation}, the transition model is validated for both turbulence models using a hypersonic  double ramp flow. For both turbulence models, the transition model is verified to predict the transition process correctly compared with experimental data. Finally, within Section \ref{sec:scram-res} the transition model is applied to three different scramjet intake configurations showing a good agreement with experimental data.

\section{Physical Modeling}\label{sec:pm}
The compressible Reynolds Averaged Navier-Stokes (RANS) equations are solved, which describe the conservation of mass, momentum and energy for compressible turbulent flows. The RANS equations read as follows:
\begin{equation}
\frac{\partial \overline{\rho}}{\partial t} + \frac{\partial}{\partial x_k}( \overline{\rho} \tilde{u}_k )=0 \;\;,
\end{equation}
\begin{equation}\label{mom_av}
\frac{\partial}{\partial t}(\overline{\rho}\tilde{u}_i) + \frac{\partial}{\partial x_k}(\overline{\rho}\tilde u_i \tilde u_k) + \frac{\partial}{\partial x_k}(\overline{\rho}\tilde R_{ik})= -\frac{\partial \overline{p}}{\partial x_i} + \frac{\partial \overline{\tau}_{ik}}{{\partial x_k}}\;\;,
\end{equation}
\begin{equation}\label{energy_av}
\frac{\partial}{\partial t} (\overline{\rho}\tilde E) + \frac{\partial}{\partial x_k}(\overline{\rho}\tilde H \tilde u_k)+\frac{\partial}{\partial x_k}(\overline{\rho}\tilde{R}_{ik}\tilde u_i) = \frac{\partial}{\partial x_k}(\overline {\tau}_{ik}\tilde u_i) - \frac{\partial \overline q_k}{\partial x_k} + \overline{\rho} D^{(k)} - \frac{\partial q^{(t)}_k}{\partial x_k}\;\;.
\end{equation}
The standard notation for the Reynolds average ($\bar{\cdot}$) and Favre average ($\tilde{\cdot} $) is employed.
The system of equations is closed using the perfect gas assumption, the Fourier assumption for the laminar and turbulent heat fluxes and the assumption of Newtonian fluid for the laminar viscous stresses. The turbulent closure is described below.

\subsection{Shear Stress Transport Turbulence Model (SST) of Menter }\label{sec:sst}
The SST model of Menter~\cite{menter:94} is a 2-equation eddy viscosity model. Like all eddy viscosity models a linear dependence between the Reynolds stress tensor and the strain rate tensor is assumed. The SST model is a combination of the $k$-$\omega$ model~\cite{wilcox:08} and the $k$-$\epsilon$ model~\cite{jones:72}. Using a blending function, in the near wall region the original $k$-$\omega$ model is used whereas further away from the wall the  $k$-$\epsilon$ model is employed. 
This was done in order to employ each model in the region where it performs best. 
The transport equations for the baseline model 
are defined as:
\begin{equation}
\frac{\partial{(\overline{\rho} k)}}{\partial{t}} 
+ \frac{\partial{(\overline{\rho} k \tilde{u}_{j})}}{\partial{x_{j}}}
 =  \tilde{R}_{ij}  \frac{\partial{u_i}}{\partial{x_{j}}}
 - \beta^{*} \overline{\rho} \omega k
 + \frac{\partial}{\partial{x_{j}}} \left[ (\mu + \sigma_{k} \mu_{t}) \frac{\partial{k}}{\partial{x_{j}}} \right ]\enspace ,
\end{equation}
\begin{equation}
\frac{\partial{(\overline{\rho} \omega)}}{\partial{t}} 
+ \frac{\partial{(\overline{\rho} \omega \tilde{u}_{j})}} {\partial{x_{j}}} 
=   \frac{\overline{\rho} \gamma_{\omega}}{\mu_{t}} \tilde{R}_{ij} \frac{\partial{u_i}}{\partial{x_{j}}}
- \overline{\rho} \beta \omega^2 
+ \frac{\partial}{\partial{x_{j}}} \left[ (\mu + \sigma_{\omega} \mu_{t}) \frac{\partial{\omega}}{\partial{x_{j}}} \right ] 
+ 2\overline{\rho}(1-F_{1}) \sigma_{\omega}  \frac{1}{\omega} \left( \frac{\partial{k}}{\partial{x_{j}}} \frac{\partial{\omega}}{\partial{x_{j}}} \right) \enspace .
\end{equation}
A detailed description of the model including all parameters can be found in \cite{menter:94}.
\subsection{SSG/LRR-$\omega$ Turbulence Model (RSM) of Eisfeld}\label{sec:ssg-lrr}
The SSG/LRR-$\omega$ model of Eisfeld~\cite{Eisfeld:06} is a combination of two previously existing models: The Speziale, Sarkar and Gatski (SSG) model~\cite{SSG} using an $\epsilon$-based length scale equation is employed in the far field and coupled to the $\omega$-based Launder, Reece and Rodi (LRR) model~\cite{LRR} in its modified Wilcox version~\cite{wi_1} for the near wall region. 
Here, the $\omega$-equation by Menter~\cite{menter:94} is employed to provide the turbulent length scale. Consequently, the blending between the two models is performed using the Menter blending function 
as well. 

The transport equations for the Reynolds stresses $ \overline{\rho}R_{ij}$ read as follows:
\begin{equation}\label{eq:rsm}
	\frac {\partial}{\partial t}(\bar{\rho}\tilde{R}_{ij} )+\frac{\partial}{\partial{x_k}}(\bar{\rho}\tilde{u}_k\tilde{R}_{ij}) = 
\bar{\rho}P_{ij}+\bar{\rho}\Pi_{ij}-\bar{\rho}\epsilon_{ij}+\bar{\rho}D_{ij}+\bar{\rho}M_{ij} \quad .
\end{equation} 

The terms on the right-hand side of the equation represent the production $\overline{\rho}P_{ij}$, the re-distribution $\overline{\rho}\Pi_{ij}$, the destruction $\overline{\rho}\epsilon_{ij}$, the diffusion $\overline{\rho}D_{ij}$, and the contribution of the turbulent mass flux $\overline{\rho}M_{ij}$, respectively. Apart from the production term, which is exact, all other terms need to be modeled. A detailed description of the model can be found in~\cite{Eisfeld:06, Bosco:11b}. 

\subsection{Transition Model: $\gamma$-$Re_{\theta_t}$ model of Langtry/Menter}\label{sec:transition-model}
The $\gamma$-$Re_{\theta_t}$ model of Langtry/Menter~\cite{menter_transition1,menter_transition2,menter_transition3} provides two additional transport equations to model the transition process. 
The $\gamma$-intermittency equation triggers the transition process and controls the production of turbulent kinetic energy in the boundary layer. The transport equation  for the transition onset Reynolds number $Re_{\theta t}$ is used to capture the non-local effect of freestream turbulence intensity and pressure gradient at the boundary layer edge. 
The reader is refered to~\cite{menter_transition3} for a detailed formulation of the model. Here, we only summarize the transport equations following the notation of Langtry/Menter:
\begin{equation}
\label{gamma_equation}
 \frac{\partial \left( \rho \gamma \right)}{\partial t} +
 \frac{\left( \rho u_{j} \gamma \right)}{\partial x_{j}} = 
 P_{\gamma} - E_{\gamma} +
 \frac{\partial}{\partial x_{j}} \left[ \left( \mu + \frac{\mu_{t}}{\sigma_{f}} \right)
                                        \frac{\partial \gamma}{\partial x_{j}} \right] \enspace ,
\end{equation}
\begin{equation}
\label{Re_theta_eqn}
 \frac{\partial \left( \rho Re_{\theta t} \right)}{\partial t} +
 \frac{\left( \rho u_{j} Re_{\theta t} \right)}{\partial x_{j}} = 
 P_{\theta t} +
 \frac{\partial}{\partial x_{j}} \left[ \sigma_{\theta t} \left( \mu + \mu_{t} \right)
                                        \frac{\partial Re_{\theta t}}{\partial x_{j}} \right] \enspace .
\end{equation}

\noindent The transition source term $P_{\gamma}$ is defined as:
\begin{equation}
\label{gamma_produktion}
 P_{\gamma} = F_{length} c_{a1} \rho S \left[ \gamma F_{onset} \right]^{0.5} \left( 1-c_{e1} \gamma \right) \enspace ,
\end{equation}

\noindent where $S$ is the strain rate magnitude,  $F_{length}$ is the transition length function, $c_{e1}=1.0$, $c_{a1}=2.0$ and $\sigma_{f}=1.0$.  
$F_{onset}$ controls the intermittency production and is defined as:
\begin{equation}
 \label{F_onset}
 F_{onset} = max \left( F_{onset2} - F_{onset3}, 0.0 \right)  \enspace ,
 \end{equation}
\begin{equation}
 \label{F_onset1}
F_{onset1} = \frac{Re_{v}}{2.193 \cdot Re_{\theta_c}}  \enspace ,
 \end{equation}
\begin{equation}
 F_{onset2} = min \left( max \left( F_{onset1}, {F_{onset1}}^4 \right) , 2.0 \right) \quad \quad , \quad \quad
\end{equation}
\begin{equation}
  F_{onset3} = max \left( 1 - \left( \frac{R_{T}}{2.5} \right)^3 , 0.0 \right) \enspace ,
\end{equation}
with
 \begin{equation}
Re_{v} = \frac{ \rho y^{2} S}{\mu} \quad \quad , \quad \quad
R_{T} = \frac{\rho k}{\mu \omega} \enspace .
\end{equation}
$Re_{\theta_c}$ in equation (\ref{F_onset1}) 
is the critical Reynolds number where the intermittency first starts to increase in the boundary layer. The empirical correlations $Re_{\theta_c}$ and $F_{length}$ can be used to calibrate the transition model to a certain flow regime (see Section \ref{sec:coupling1} - \ref{sec:coupling3}). For all other variables the reader is refered to~\cite{menter_transition3}. For a better prediction of separation-induced transition, the effective intermittency $\gamma_{\mathit{eff}}$ as defined by Langtry/Menter~\cite{menter_transition3} is used to modify the source terms of the turbulence model.

\subsubsection{Coupling to the SST Turbulence Model by Langtry/Menter}\label{sec:coupling1}
The transition model was originally designed to be coupled with the SST model. Here, we follow the approach of Langtry/Menter. Only minor changes in the production and destruction term of the $k$-equation have to be performed. The equation for the specific turbulence dissipation rate $\omega$ is unaltered, whereas the production term of the $k$-equation is modified as follows:
\begin{equation}
P_{k_{mod}} = \gamma_{\mathit{eff}} \cdot P_{k_{orig}} \quad \quad , \quad \quad
D_{k_{mod}} = min \left( max \left( \gamma_{\mathit{eff}} , 0.1 \right) , 1.0 \right) \cdot D_{k_{orig}} \enspace .
\end{equation}
Some final modifications have to be done for the blending function $F_1$ of the original SST model and these are defined as:
\begin{equation}\label{eq:F1mod}
 F_{1_{mod}} = max \left( F_{1_{orig}} , F_3 \right) \quad , \quad \quad
F_3 = e^{ - \left( \frac{R_y}{120.0} \right)^8} \quad \quad , \quad \quad 
R_y = \frac{\rho y \sqrt{k}}{\mu} \enspace .
\end{equation}

The following correlations were developed by Langtry/Menter for subsonic and transonic flows:
 \begin{equation}
Re_{\theta_c}  = \left\{
\begin{array}{cl}
\overline{Re_{\theta_t}}
-396.035\cdot 10^{-2} 
 +120.656\cdot 10^{-4}\overline{Re_{\theta_t}}  
-868.230\cdot 10^{-6})\overline{Re_{\theta_t}}^2& \\
+696.506\cdot 10^{-9})\overline{Re_{\theta_t}}^3
-174.105\cdot 10^{-12})\overline{Re_{\theta_t}}^4 \quad ,\quad \quad &\overline{Re_{\theta_t}}\leq 1870\\
\overline{Re_{\theta_t}} - \left(593.11 +\left( \overline{Re_{\theta_t}} + 1870.0\right) \cdot 0.482\right) \quad , \quad \quad &\overline{Re_{\theta_t}}> 1870
\end{array}
\right.
\end{equation}

\begin{equation}
F_{length}= \left\{
\begin{array}{cl}
 98.189 \cdot 10^{-1} - 119.270 \cdot 10^{-4} \overline{Re_{\theta_t}} -132.567\cdot 10^{-6} \overline{Re_{\theta_t}}^{2} \quad ,\quad   &\overline{Re_{\theta_t}}< 400\\
263.404 - 123.939 \cdot 10^{-2} \overline{Re_{\theta_t}} +194.548\cdot 10^{-5} \overline{Re_{\theta_t}}^{2} &\\ -101.695\cdot 10^{-8} \overline{Re_{\theta_t}}^{3}  \quad ,\quad  &400\leq \overline{Re_{\theta_t}}< 596\\
 0.5 - \left(\overline{Re_{\theta_t}} -596.0\right)\cdot 3.0\cdot 10^{-4}  \quad ,\quad   &596 \leq  \overline{Re_{\theta_t}}< 1200 \\
0.3188 \quad .\quad   & 1200 \leq \overline{Re_{\theta_t}}
\end{array}
\right.
\end{equation}

\subsubsection{Coupling to the SST Turbulence Model by Krause}\label{sec:coupling2}
The correlations by Langtry/Menter were tested for hypersonic flows predicting a wrong transition location. Therefore hypersonic, in-house correlations~\cite{krause:10} were developed. The modifications of the turbulence equations are unchanged.

Langtry/Menter~\cite{menter_transition3} proposed the correlations $Re_{\theta_c}=f(\overline{Re_{\theta_t}})$ and $F_{length}=f(\overline{Re_{\theta_t}})$ for subsonic and transonic speed. However, for hypersonic flows, $\overline{Re_{\theta_t}}$ is much higher than for subsonic flows. In hypersonic regimes, $\overline{Re_{\theta_t}}$ can easily be in the order of $10^5$, e.g., in reattachment zones, whereas for subsonic regimes it is more than 50 times smaller. Therefore our correlations depend on the freestream turbulent intensity $I_\infty$ instead of $\overline{Re_{\theta_t}}$:
\begin{equation}
 \label{Re_theta_c}
Re_{\theta_c} = 967.34 \cdot I_\infty^{-1.0315} \quad \quad , \quad \quad
F_{length} = 10.435 \cdot I_\infty^{2.9756} \enspace .
\end{equation}
Hence, $\overline{Re_{\theta_t}}$ is only used in the modification for separation-induced transition to compute $F_\theta$. Therefore, in principle the second transport equation can be disregarded when finding a different modification for separation-induced transition.

\subsubsection{Coupling to the RSM Turbulence Model}\label{sec:coupling3}
%
Since the LRR/SGG-$\omega$ model uses the same $\omega$-equation as the SST model, we propose to change the source terms of the Reynolds stress transport equations~\eqref{eq:rsm} in a similar manner. The equation for the specific turbulence dissipation rate $\omega$ is unaltered. The source terms of the Reynolds stress transport equations are modified as follows:
\begin{equation}
P_{ij_{mod}} = \gamma_{\mathit{eff}} \cdot P_{ij_{orig}} \quad \quad, \quad \quad
\Pi_{ij_{mod}} = \gamma_{\mathit{eff}} \cdot \Pi_{ij_{orig}} \enspace , 
\end{equation}
\begin{equation}
D_{ij_{mod}} = min \left( max \left( \gamma_{\mathit{eff}} , 0.1 \right) , 1.0 \right) \cdot D_{ij_{orig}} \enspace .
\end{equation}
The modification for the blending function $F_1$ of the original SST model is given in (\ref{eq:F1mod}).
The following in-house correlations are used for hypersonic flows:
\begin{equation} \label{eq:corr-RSM}
Re_{\theta_c} = 949.6376 \cdot I_\infty^{-0.5379}-254.9323 \enspace ,  
\end{equation}
\begin{equation} \label{eq:corr2-RSM}
F_{length} = 0.0045 I_\infty - 0.0902  I_\infty^2 +0.2343  I_\infty^3 +1.2776  I_\infty^4\enspace .
\end{equation}
Details on these modifications are given in Section~\ref{sec:mod-trans-rsm}. 
\section{Numerical Methods}\label{sec:qf}

\subsection{QUADFLOW Solver}
The in-house program QUADFLOW has been extensively validated over the last two decades~\cite{Bramkamp:03b,Nguyen:11,Krause:07,Bosco:11b}.
The program solves the RANS equations for unsteady, compressible fluid flow in two and three dimensions using a cell-centered finite volume discretization \cite{Bramkamp:04}.

Mesh-adaptation is realized by wavelet-based multiscale techniques \cite{Mueller:03,mueller:09}. 
Starting point for this adaptation procedure is a hierarchy of nested grids $G_l:=\{\Omega_{l,i}\}_{i\epsilon {\cal I}_l}$, $l=0,...,L$ and corresponding averages $\bar{u}_l=\{u_{l,i}\}_{i\epsilon {\cal I}_l}$ for all variables by which grid adaptation is performed (e.g., all mean flow variables). By means of this hierarchy, the averages $\bar{u}_L$ on the finest level $L$ are successively decomposed into a sequence of averages on the coarsest level $\bar{u}_0$ and details $\bar{d}_l$ ($l=0,...,L$). The details $\bar{d}_l$ describe the local update of the solution on two successive refinement levels. They can be interpreted as differences, which become negligibly small in regions, where the solution is sufficiently smooth. Thus, a set of significant details can be defined $D_{\varepsilon} :=\{(l,i):\left|d_{l,i}\right| > \varepsilon_l\}$, where $\varepsilon_l = 2^{l-L}\varepsilon$ is a level-dependent threshold value. During the adaptation procedure all cells with significant details are refined. The threshold value $\epsilon$ is set by the user and determines the sensitivity of the grid adaptation~\cite{Mueller:03,mueller:09}, i.e., more cells are refined for deceasing threshold value.

For steady state problems, the computations start on the uniform level $l=1$ grid. Mesh adaptation is performed whenever the averaged density residual drops below a certain user-defined value $Res_{drop}$. After the last adaptation, the simulation continues until the steady state is reached.

The mesh is treated as fully unstructured and composed of polygonal (2D) or polyhedral (3D) elements. It is based on a multi-block B-spline representation \cite{SM-Lamby:07}. This approach is especially suited for dealing with hanging nodes appearing in locally refined meshes. For the time and space discretization, the user can choose among several options for the Riemann solver, the limiter, the reconstruction and the time integration. Here, we summarize the methods used for the computations presented in this paper:
The convective fluxes are discretized using the AUSMDV Riemann solver \cite{wang:98}. A linear Green-Gauss reconstruction \cite{barth:89} of the primitive variables is performed to locally achieve second-order accuracy in space, and the Venkatakrishnan slope limiter is employed to avoid oscillations typical of higher order schemes \cite{limiter}. For the discretization of the viscous fluxes, a modified central difference method is used \cite{Bramkamp:03}. 
A second order accurate explicit Runge-Kutta scheme \cite{swanson:85} is employed for the time integration using a maximum CFL number of 3.0. For the treatment of turbulent flows, the SST model, the SST-Tr model, the RSM model and the RSM-Tr model are used.
The parameters to control the adaptation are $\varepsilon_{thres}=10^{-3}$ and $Res_{drop}=10^{-2}$ for all test cases.
This solver has been parallelized on distributed memory architectures using MPI \cite{Brix:09,Brix:11}. For load-balancing, we use the concept of space-filling curves \cite{Zumbusch:03}. Just recently this approach was successfully applied to hypersonic applications including fully three-dimensional computations of scramjet intakes \cite{Frauholz:14c}. 

\subsection{Boundary Conditions}
\paragraph{Conservative Quantities}
At the far field boundaries, supersonic inflow or outflow conditions are imposed. At solid boundaries, the no-slip condition and an isothermal wall are prescribed. For three-dimensional simulations of the intake, a half model is used and a symmetry condition is imposed at one side.

\paragraph{Turbulent Quantities} 
The turbulent values are determined by the freestream turbulence intensity $I_\infty$: $k_\infty = 1.5(I_\infty u_\infty)^2$. The Reynolds stress matrix is initialized by placing $2/3 k_\infty$ on the diagonal and the
specific dissipation rate of the freestream is $\omega_\infty = k_\infty/(RLTU \cdot \mu_l)$ with RLTU=0.001 being a measure for the ratio of turbulent to laminar viscosity in the freestream. 
The no-slip condition also implies that the Reynolds stresses are zero at the wall. The chosen $\omega$-wall condition is the one from Menter \cite{menter:94} imposing a value of this quantity depending on the distance of the first cell center from the wall. 
 
\paragraph{Transitional Quantities}
The freestream values of $\gamma=1$ and $\overline{Re_{\theta_t}}=0.0001$ are used at the inflow boundary. For the supersonic outflow, the variables are extrapolated from the interior. At no-slip walls, $\gamma$ is set to zero and a value $\overline{Re_{\theta_t}}=0.0001$ is employed.

\section{Modifications of the hypersonic correlations for the RSM-Tr model}\label{sec:mod-trans-rsm}
For the RSM-Tr model, the hypersonic in-house correlations developed for the SST model result in a wrong prediction of the transition location showing the necessity of calibrating the correlations for the more sensitive RSM-Tr model. 
For the calibration process, we consider a hypersonic flat plate at M$_\infty=6.3$ with different freestream enthalpies and Reynolds numbers. This test case was also used to develop the hypersonic in-house correlations for the SST model coupled to the transition model (SST-Tr).
For the RSM-Tr, first we tested the ansatz used for the SST-Tr model:
\begin{equation}\label{eq:corr-sst-ansatz}
Re_{\theta_c}=A\cdot I_\infty^{-B}\quad \quad , \quad \quad
F_{length}=C\cdot I_\infty^{D} \enspace .
\end{equation}
Since the correlations depend only on the freestream turbulence intensity, the correlations reduce to constant values. 
We find optimal values for each of the four test conditions by changing $Re_{\theta_c}$ and $F_{length}$ manually (see Tab. \ref{tab:re-fl}). Then we determine the coefficients within the correlations by solving a least squares problem.
\begin{table*}[!ht]
\begin{tabular}{c|c|c|c}
condition 	& $I_{\infty}$	& $Re_{\theta_t}$  	& $F_{length}$   \\ \hline
   1      		& 5.7\%					& 120				 	 	& 1390      \\ 
   2 			& 4.4\%					& 170					 	& 500      \\ 
   3 			& 3.7\%					& 215  						& 250      \\ 
   4 			& 3.8\%					& 210						& 380    
   \end{tabular}
 \caption{Optimal values for the transition onset Reynolds number and the transition length function.}\label{tab:re-fl}
 \end{table*}

However, the resulting correlations using ansatz \eqref{eq:corr-sst-ansatz} did not predict correctly the transition process for the validation test case. Hence, we modified the ansatz as follows:
\begin{equation}\label{eq:corr-base}
Re_{\theta_c}=A\cdot I_\infty^{-B} - C\quad \quad , \quad \quad
F_{length}=D\cdot I_\infty+E\cdot I_\infty^2+F\cdot I_\infty^3+G\cdot I_\infty^4\enspace .
\end{equation}

\begin{figure}[ht!]
  \includegraphics[width=0.49\textwidth,clip]{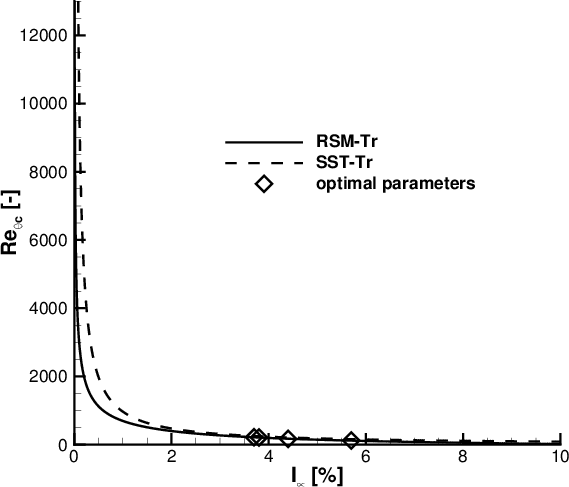}
    \includegraphics[width=0.49\textwidth,clip]{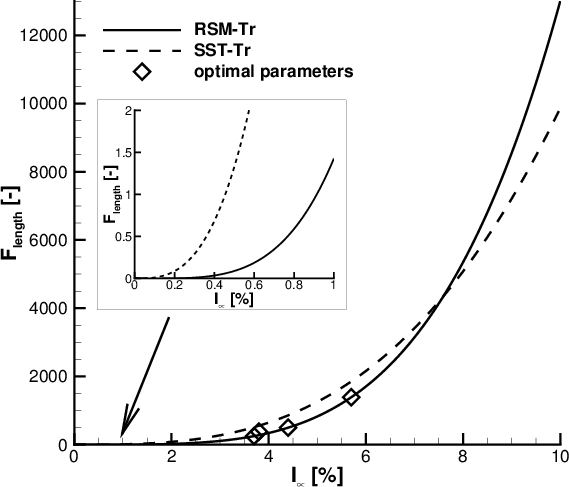}
  \caption{Correlation functions $Re_{\theta_c}$ (left) and $F_{length}$ (right) for the RSM-Tr model.}
\label{fig:Korrelation}
\end{figure}
Figure \ref{fig:Korrelation} shows the two correlation functions and the optimal values for the RSM-Tr model. The correlations of the SST-Tr model are shown as well.
Note that the freestream turbulence intensity $I_\infty$ within the correlations is a unique parameter of the experimental facilities and to some extent of the investigated flow conditions. Often, this parameter is not known from the experiments. In this case, one can determine the turbulence intensity numerically using one experiment and then apply the same turbulence intensity for all other computations for this facility and the corresponding flow condition. Within this work, we are using the  freestream turbulence intensity determined numerically by Krause~\cite{krause:10}.

In the next section, the influence of the correlations is investigated to illustrate the calibration process.

\subsection{Hypersonic Flat Plate}
The considered flat plate is 1.5~m long and 0.12~m wide. Experiments were done by Mee \cite{mee:02} in a T4 piston-free shock tunnel without tripping the boundary layer. Flush-mounted thin-film heat-transfer gauges were used to detect the location of transition.  The inflow conditions used for the numerical computations are summarized in Table \ref{tab:HFP_inflow}. 
\begin{table}[ht!]
 \begin{center}
  \begin{tabular}{|c|c|c|c|c|}\hline
                     & low enthalpy        & low enthalpy        & low enthalpy         & high enthalpy \\
 condition           & low Re              & mid Re              & high Re              & low Re \\
                     & condition 1         & condition 2        & condition 3   & condition 4 \\\hline
 nozzle enthalpy [MJ/kg]& 5.3              & 6.2                 & 6.8                  & 12.4 \\ \hline
 $M_\infty$ [-]      & 6.3                 & 6.2                 & 6.1                  & 5.5 \\ \hline
 $T_\infty$ [K]      & 570                 & 690                 & 800                  & 1560 \\ \hline
 $Re_\infty$ [1/m]   & $1.7\cdot 10^6$     & $2.6\cdot 10^6$     & $4.9\cdot 10^6$      & $1.6\cdot 10^6$ \\ \hline
 $I_\infty$ [\%]    & 5.7             & 4.4                 & 3.7                  & 3.8     \\ \hline
  \end{tabular}
 \end{center}  \caption{Test conditions for the flat plate.} \label{tab:HFP_inflow}
\end{table}


The grid contains 16 cells in the flow direction and 6 cells in the cross-flow direction at level $l=0$. Cells are clustered near the leading edge and toward the solid wall to obtain the desired resolution of 10$^{-6}$~m on the finest level $l=4$ in these regions. The final level $l=4$ grid contains approximately 17,000 cells. 

\begin{figure}[t!h]
  \includegraphics[width=0.49\textwidth, ,clip]{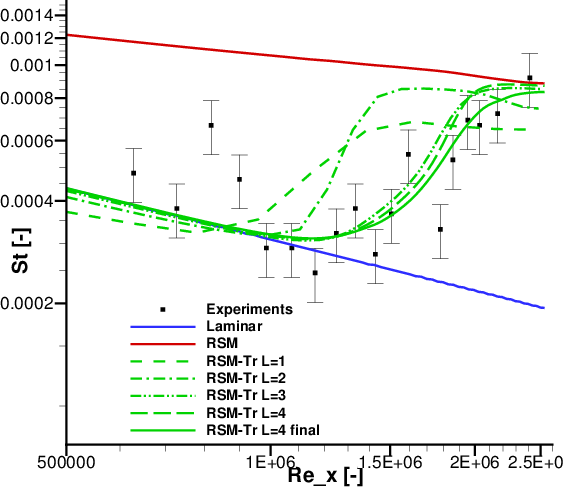}
    \includegraphics[width=0.49\textwidth, clip]{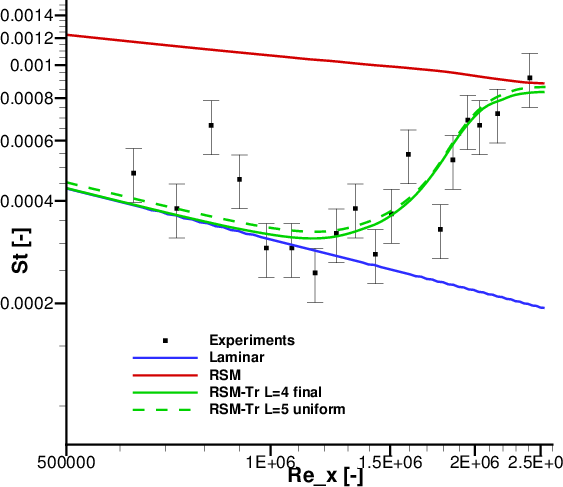}
  \caption{Influence of grid adaptation on the transition onset location and the transition length for condition 1 by means of the Stanton number distribution.}
\label{fig:HFP_GridConv}
\end{figure}
Figure \ref{fig:HFP_GridConv} (left) shows the Stanton number,
\begin{equation}
St = \frac{q_w}{\rho_{\infty} \left| u_{\infty} \right| c_p (T_{0,\infty}-T_w)} \enspace ,
\label{eq:stanton}\end{equation}
\noindent for the different levels of the adaptive computations. Here the huge impact of the first wall distance on the transition onset point and the transition length can be seen. For larger first wall distances, the transition onset point moves upstream and the transition length is increased. To prove grid convergence of the results, the adaptive solution at level $l=4$ is compared to the solution of the uniformly refined grid on the next level $l=5$. As Figure \ref{fig:HFP_GridConv} (right) shows the results can be considered grid-converged.

\begin{figure}[ht!]
  \includegraphics[width=0.49\textwidth, clip]{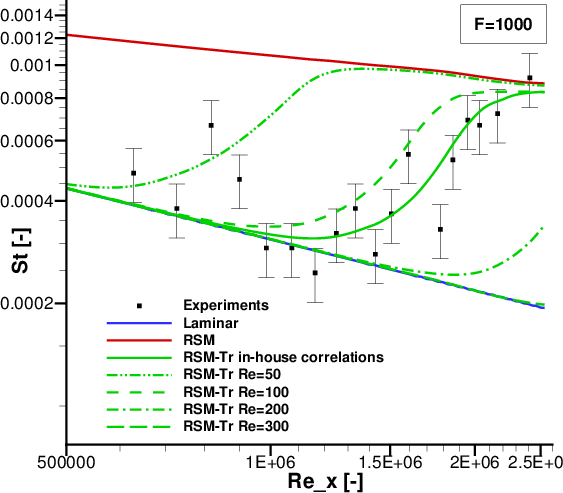}
  \includegraphics[width=0.49\textwidth, clip]{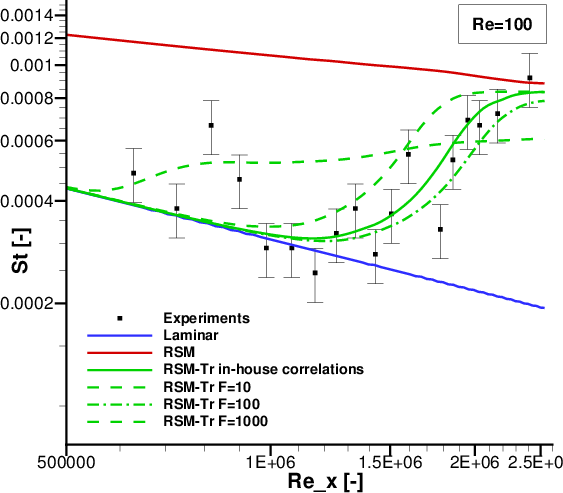}
  \caption{Influence of the transition onset Reynolds number (left) and the transition length function (right) for condition 1 by means of the Stanton number.}
\label{fig:HFP_DevCorr1}
\end{figure}
To investigate the influence of the transition onset Reynolds number $Re_{\theta_c}$  and the transition length function $F_{length}$, computations with variable $Re_{\theta_c}$ for constant $F_{length}=1000$ as well as computations with variable F$_{length}$ for constant $Re_{\theta_c}=100$ are performed for condition 1. The Stanton number distribution is shown in Figure \ref{fig:HFP_DevCorr1} for different values of $Re_{\theta_c}$ (left) and for different values of $F_{length}$ (right).
Larger values of $Re_{\theta_c}$ result in a more laminar solution. The transition onset point moves  further downstream whereas the transition length is not influenced. The value of $F_{length}$ has an impact on both, the transition onset point and the transition length. The transition length increases for decreasing values of $F_{length}$. Also, smaller values of $F_{length}$ lead to an earlier transition onset point. The same was found for the SST-Tr model for subsonic and supersonic test cases~\cite{krause:10}.

\begin{figure}[ht!]
  \includegraphics[width=0.49\textwidth, clip]{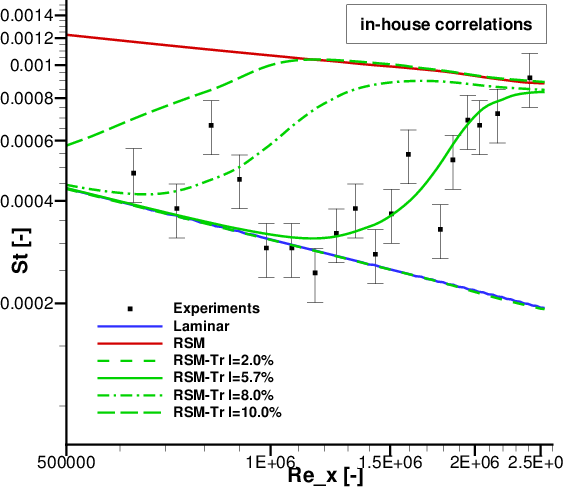}
\includegraphics[width=0.49\textwidth, ,clip]{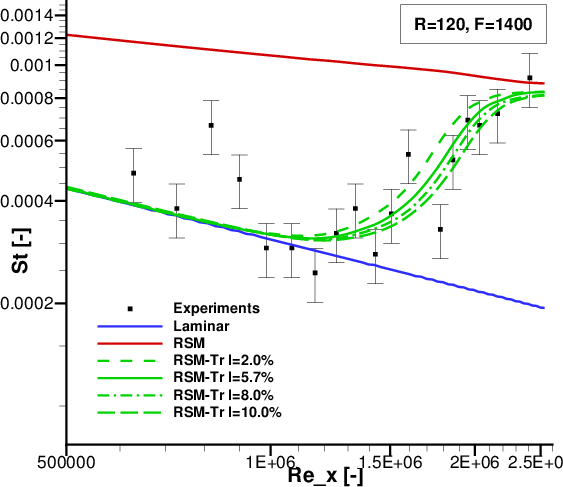}
  \caption{Influence of freestream turbulence intensity for the in-house correlations (left) and fixed values of $Re_{\theta_c}$ and $F_{length}$ (right) for condition 1 by means of the Stanton number distribution.}
\label{fig:HFP_DevCorr2}
\end{figure}

It is obvious, that due to the chosen correlations the freestream turbulence intensity strongly impacts the numerical solution (Fig. \ref{fig:HFP_DevCorr2} left). An additional small impact of the freestream turbulence intensity on the transition process can be found when varying $I_\infty$ for constant values of $Re_{\theta_c}$ and $F_{length}$. This is illustrated in Figure \ref{fig:HFP_DevCorr2} (right) showing the Stanton number distribution. The transition onset point and the transition length are not affected, but the curvature of the profile is. For smaller $I_\infty$, the Stanton number increases faster close to the transition onset point and decreases faster at the transition end point.

\begin{figure}[ht!]
  \includegraphics[width=0.49\textwidth, clip]{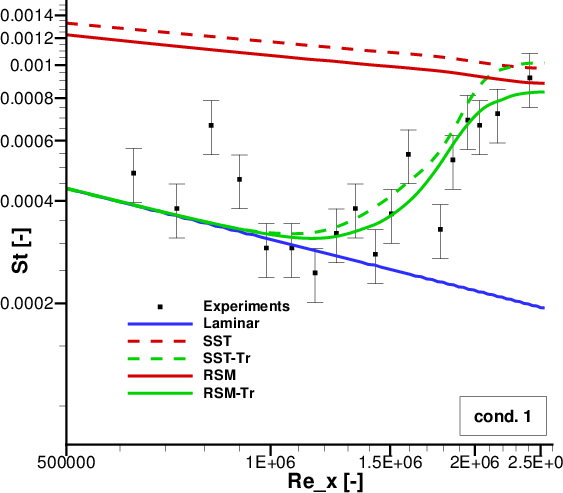}
  \includegraphics[width=0.49\textwidth,clip]{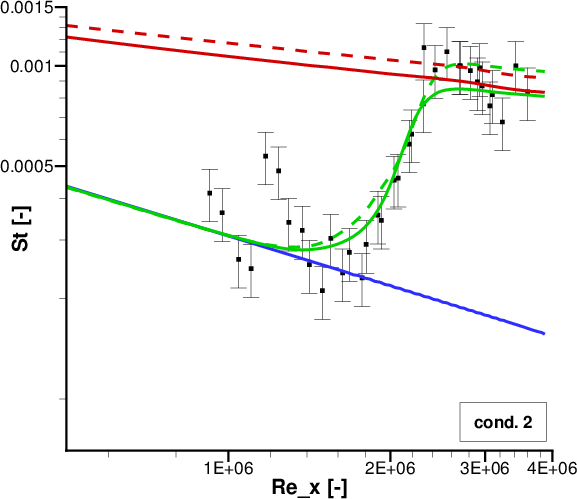}
  \includegraphics[width=0.49\textwidth, clip]{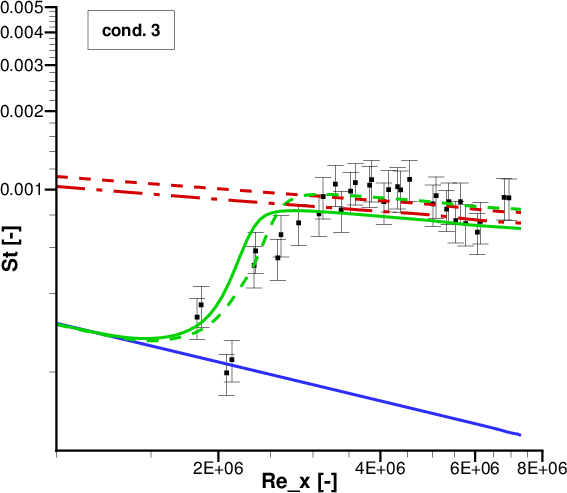}
  \includegraphics[width=0.49\textwidth, clip]{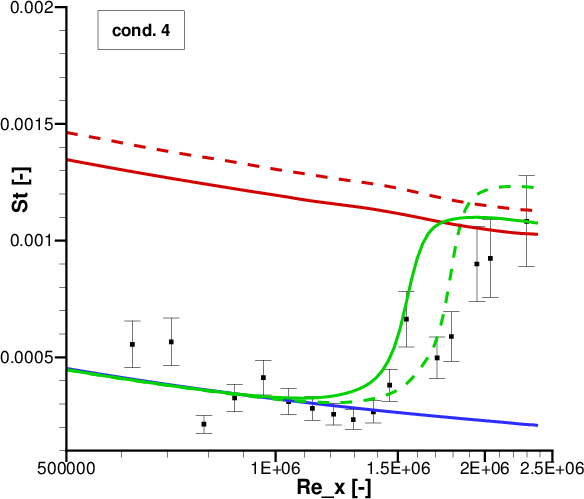}
  \caption{Stanton number distribution for different test conditions.}
  \label{fig:HFP_Kond34}
\end{figure}

Figure \ref{fig:HFP_Kond34} shows the Stanton number for the SST-Tr model, the RSM-Tr model using the newly developed in-house correlations and the experiments for all test conditions. 
Overall, the agreement between the numerical results and the experimental data is good. The transition onset point and the transition length are detected correctly in all cases. The RSM-Tr model predicts the same transition onset point but a slightly higher transition length than the SST-Tr model. From the experimental data, it is not possible to decide which model performs best.
\newpage
%
\section{Validation for the RSM-Tr model}\label{sec:validation}
This section validates the RSM-Tr model using a double ramp test case at $M_\infty=8.1$. Experimental data are shown as well. Experiments for the double ramp were performed at the hypersonic shock tunnel TH2 of the Shock Wave Laboratory (SWL) at RWTH Aachen University~\cite{neuenhahn:06,neuenhahn:09}. During the campaign the influence of 
the wall temperature on the flow field with focus on the separation bubble and laminar-to-turbulent transition was investigated. The wall pressure and heat flux were measured via Kulite piezo-resistive sensors and type K coaxial thermocouples, respectively. The inflow conditions are given in Table \ref{tab:HDR-inflow}.
The ramp angle for the first and second ramp are 9 degrees and 20.5 degrees, respectively. The first ramp length is 180~mm and the second ramp length is 255~mm, both measured along the ramp surface. The model is 270~mm wide to ensure two-dimensional flow.
\begin{table*}[!ht]
\begin{center} 
   \begin{tabular}{|c|c|c|c|c|c|} \hline
      $M_\infty$ [-] & $Re_\infty$ [$10^6$/m] &   $T_0$ [K]&   $T_\infty$ [K] &   $T_{\rm w}$ [K] &$I_\infty$ [\%]\\\hline
            8.1  &  3.8            	     &  1635       &     106         &    [300, 600, 760] 	&0.9	\\\hline
 \end{tabular}
 \end{center} 
 \caption{Test conditions for the double ramp.}\label{tab:HDR-inflow}
 \end{table*}

\begin{figure*}[!ht]
 \centerline{\includegraphics [width = 0.9\textwidth,clip]{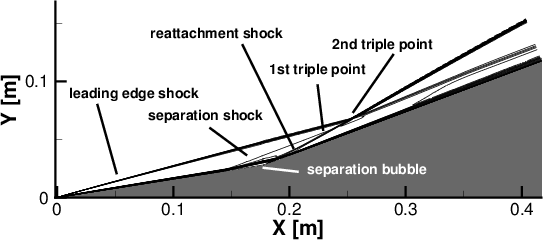}}
 \caption{Computed Mach number lines for the double ramp.} \label{fig:HDR-sharple-300K-M}
  \end{figure*}
The general flow phenomena are similar for all investigated wall temperatures. Figure \ref{fig:HDR-sharple-300K-M} shows the flow features for the lowest wall temperature 300~K. 
The leading edge shock is attached and a separation bubble occurs at the end of the first ramp. At the first triple point, the separation shock and the reattachment shock merge into one shock wave. At the second triple point, the resulting shock and the leading edge shock are merged and form a single shock wave.

\begin{figure*}[!ht]
 \centerline{\includegraphics [width = 0.9\textwidth,clip]{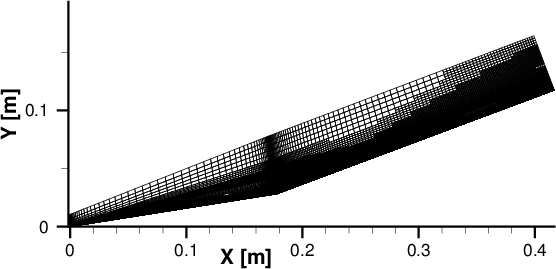}}
\caption{Final grid at refinement level $l=4$ for the double ramp for wall temperature 300~K .}\label{fig:HDR-grid}
 \end{figure*}

The grid has 16 cells in flow direction along the first ramp and 16 cells in flow direction along the second ramp at level $l=0$. Perpendicular to the flow, the grid has 8 cells at level $l=0$. To ensure a first wall distance of 10$^{-6}$~m on the final level $l=4$, the cells are clustered towards the walls. To prove grid convergence, the results of the adaptive computations are compared to results obtained on a uniform refined grid at $l=5$ showing no significant difference (not shown here). Thus, the results are considered to be grid-converged. The final adapted grid containing 45,000~cells is shown in Figure \ref{fig:HDR-grid}.

\begin{figure}[ht!]
  \includegraphics[width=0.49\textwidth,clip]{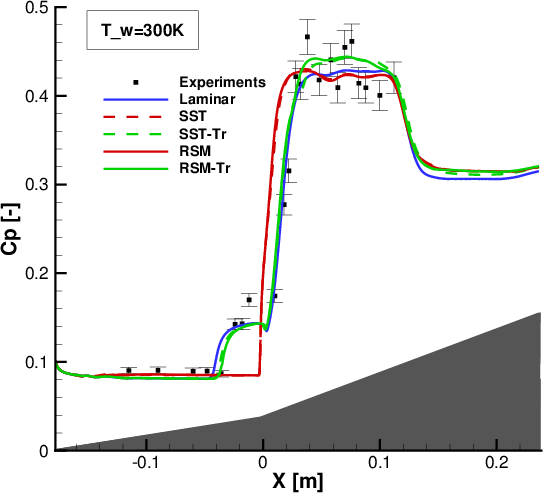}
  \includegraphics[width=0.49\textwidth,clip]{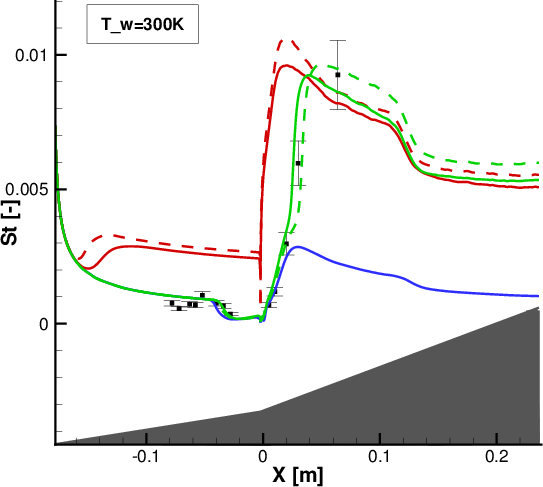}
  \includegraphics[width=0.49\textwidth,clip]{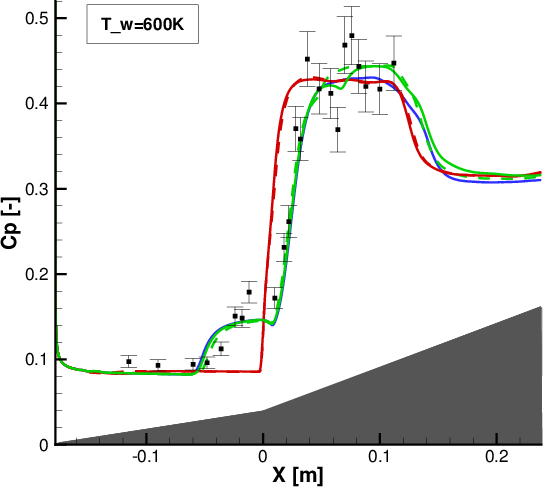}
  \includegraphics[width=0.49\textwidth,clip]{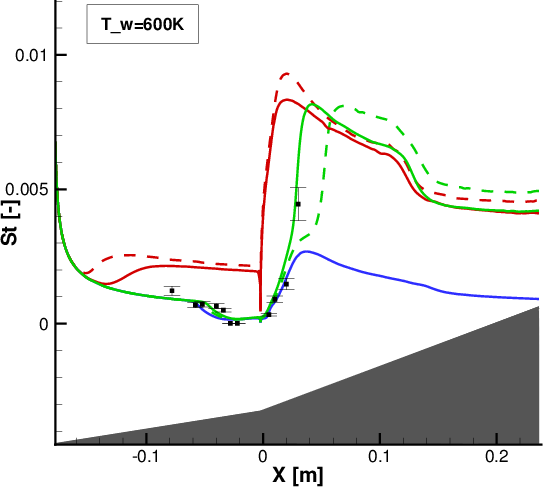}
  \includegraphics[width=0.49\textwidth,clip]{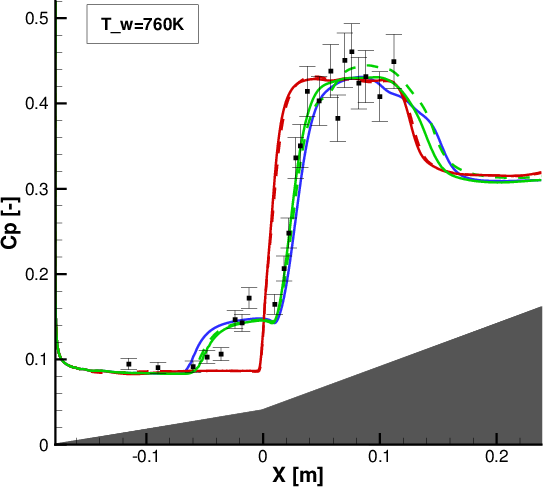}
  \includegraphics[width=0.49\textwidth,clip]{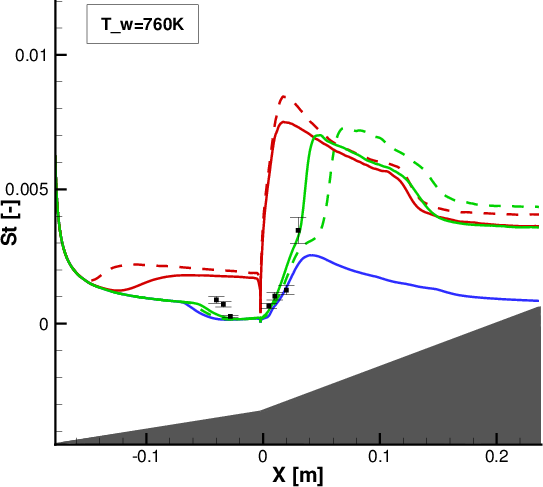}
  \caption{Pressure coefficient (left) and Stanton number distribution (right) for the double ramp with sharp leading edge at different wall temperatures.}
  \label{fig:HDR-cp-st}\end{figure}
Figure \ref{fig:HDR-cp-st} (left) shows the pressure coefficient for the different wall temperatures. 
Due to the fully laminar boundary layer, the separation bubble predicted by the laminar computation is larger than the separation bubble predicted by the transitional computations. The fully turbulent computations do not predict a separation at all due to the physics of turbulent boundary layers. At the first ramp, the pressure loads increase due to the separation shock wave. This is followed by a plateau in  the pressure over the separation bubble and a pressure increase caused by the reattachment shock. The next higher pressure plateau starting at the first triple point, and the lower pressure plateau starting at the second triple point are captured correctly. 
After the second triple point, the pressure level is lower since the flow passes only one merged shock. Compared to experimental data the transitional computations show a very good overall agreement. The separation predicted by the RSM-Tr model is slightly smaller than the separation predicted by the SST-Tr model and fits better to the experiments. For wall temperature 760~K, the separation size is slightly overpredicted by both transitional computations.

In Figure \ref{fig:HDR-cp-st} (right), the Stanton number is presented for all wall temperatures. The laminar separation predicted by the laminar and transitional computations can be seen in the drop of the Stanton number. The triple points are also visible near a higher and a lower heat load plateau. For the transitional computations the boundary layer transitions over the separation bubble from laminar to turbulent and, thus, the heat loads along the second ramp are at the turbulent level. The overall agreement of both transitional computations with the experimental data is very good. The RSM-Tr model predicts the transition onset point slightly earlier than the SST-Tr model.
The computations verify the capability of the transition model to predict the effects of changing wall temperature. 

\section{Results: 3D Scramjet Intake Flows}\label{sec:scram-res}
In the next three subsections, results for three different scramjet intake configuration are discussed using the RSM-Tr model and SST-Tr model. Usually we start numerical investigation of scramjet intakes with applying the SST model and SST-Tr model. Depending on the agreement to experimental data, we continue the analysis using the computationally more expensive and numerically less stable RSM model and RSM-Tr model. 
Hence, for the first intake the RSM model and RSM-Tr model are compared. Computations using the SST model and SST-Tr model can be found in~\cite{Nguyen:10,Nguyen:11}. The second intake is computed using the SST model and the SST-Tr model showing a good agreement to experimental measurements. Since no experimental data is available yet for the third intake, computations with SST model, SST-Tr model, RSM model and RSM-Tr model are performed.  

\subsection{SWL Intake}\label{Sec:SWL}
\begin{figure}[!ht]
\begin{center}
  \includegraphics[height=0.3\textwidth,clip]{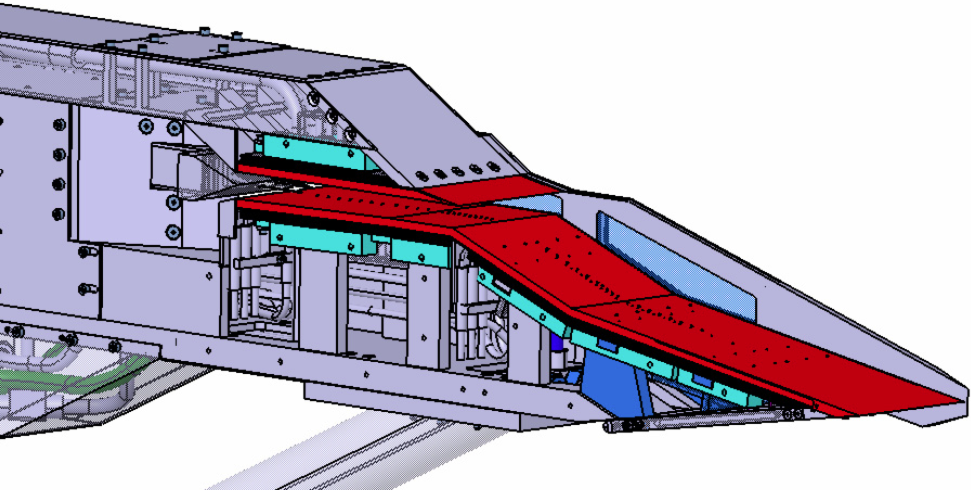}
  \caption{CAD model of the SWL intake. Reproduced from \cite{Fischer}.}
  \label{geo}
\end{center}
\end{figure}

The intake model considered within this section has been developed in the frame of the German Research Training Group GRK 1095 ``Aero-Thermodynamic Design of a Scramjet Engine for Future Space Transportation Systems'' \cite{Gaisbauer:07} and was built and tested at the Shock Wave Laboratory (SWL) at RWTH Aachen University. Figure \ref{geo} shows the geometry of the considered scramjet intake. The model has two exterior compression ramps and an interior section. The leading edge of the first ramp and the cowl lip are sharp. The model is 100 mm wide and has straight side walls on both sides. The side wall has a starting height of 9~mm and an angle of 16 degrees.

The configuration has been designed for an inflow Mach number $M_\infty=7.5$ and was tested at a slight off-design condition in the hypersonic shock tunnel facility TH2 in Aachen \cite{neuenhahn:06, Fischer}. The test conditions of the experimental campaign are listed in Table \ref{t:condi}. These values are used as inflow conditions in the simulations. During the experiments, pressure and heat transfer rate were measured by Kulite pressure probes and thermocouples, respectively.
\begin{table}[!ht]
\begin{tabular}{|c|c|c|c|c|c|} \hline
      $M_\infty$ [-] & $Re_\infty$ [$10^6$/m] &   $T_0$ [K]&   $T_\infty$ [K] &   $T_{\rm w}$ [K] &$I_\infty$ [\%]\\\hline
            7.7  &  4.1            	     &  1520       &     125         &    300 	&0.9	\\\hline
 \end{tabular}
 \caption{\label{t:condi}Test conditions for the SWL scramjet intake configuration.}
\end{table}

For the numerical analysis, the grid has 44 cells in the flow direction, 6 cells in the cross-flow direction and 5 cells in spanwise-direction on refinement level $l=0$. Only the half-model is computed using a symmetry boundary condition.
To ensure a minimum wall distance of $10^{-6} $~m on the final level $l=4$ grid, the grid points in wall-normal direction are stretched towards the walls using a Poisson distribution. Transverse to the wall, the grid lines are almost always orthogonal to the walls to resolve accurately the strong wall gradients. The final adapted grid on level $l=4$ has 3.5 million cells. Grid convergence was shown in \cite{Frauholz:14c} using a fixed transition point at the end of first ramp. 

\begin{figure}[!ht]
\begin{center}
    \includegraphics[width=1\textwidth,clip]{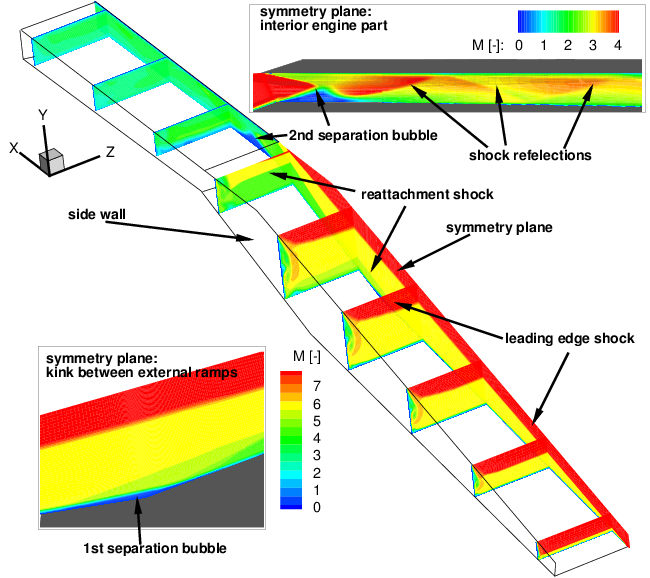}
  \caption{Computed Mach number distribution for the half-model of the SWL intake (RSM-Tr model).}
  \label{fig:swl:3d-M}
\end{center}
\end{figure}
The overall flow phenomena can be seen in Figure \ref{fig:swl:3d-M}. The incoming flow is first compressed through an oblique shock wave generated by the sharp leading edge. Additional compression is achieved through an oblique shock wave produced by the side walls. A laminar boundary layer develops along the first ramp. Between the first and second compression ramp the flow separates and transitions from laminar to turbulent. 
Since the intake was tested at off-design conditions, the reattachment shock hits the upper intake wall and deflects the oblique shock wave produced by the cowl lip slightly.  Due to the impingement of the cowl shock on the expanding flow and ramp boundary layer, a second flow separation on the lower engine wall is produced. The cowl shock is reflected several times at the upper and lower engine wall in the interior region.

   \begin{figure*}[!ht]
 \begin{center}
 \includegraphics [width = 1\textwidth,clip]{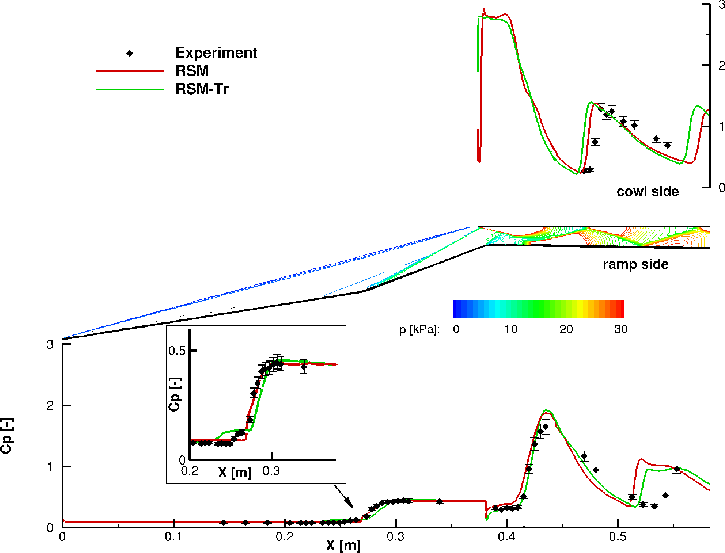}
  \end{center}
        \caption{Pressure coefficient distribution of the scramjet intake in the symmetry plane.}\label{swl-cp-wall-scram}
          \end{figure*}
          \begin{figure*}[!ht]
 \begin{center}
  \includegraphics [width = 1\textwidth, clip]{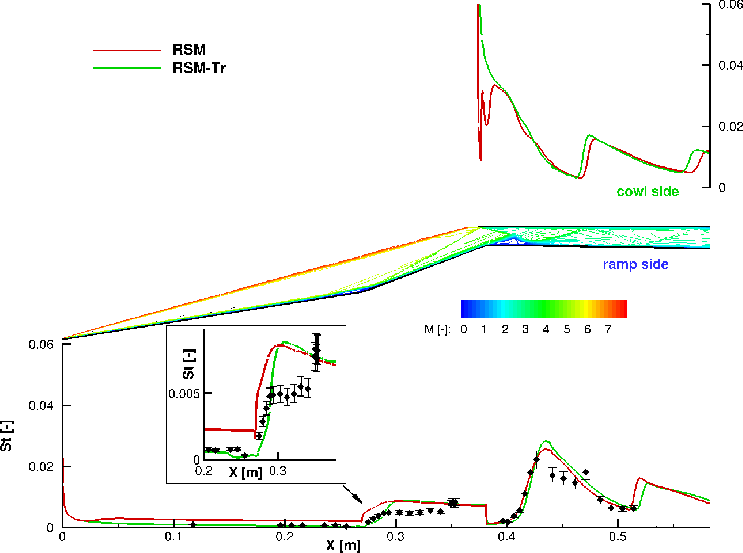} 
  \end{center}
        \caption{Stanton number distribution of the scramjet intake in the symmetry plane.}\label{swl-st-wall-scram}
  \end{figure*}  
A comparison of the RSM-Tr computation to experimental data \cite{neuenhahn:06, Fischer} is shown in Figure \ref{swl-cp-wall-scram} and \ref{swl-st-wall-scram} for the pressure coefficient and the Stanton number along the centerline. The fully turbulent result of the RSM computation is shown for reference. The flow predicted by the RSM-Tr model starts laminar at the first ramp and follows the experimental data closely. The size of the separation bubble between the two compression ramps is slightly overpredicted by the RSM-Tr computation. Here, the flow becomes transitional within the shear layer. The measured Stanton numbers after reattachment indicate that the transition point is predicted slightly too early by the RSM-Tr model. Compared to fully turbulent RSM simulation not predicting the separation, the agreement with experimental data is improved. The impingement of the cowl shock wave from the upper engine wall causes a second separation bubble. In the separated flow area and the subsequent reattachment peak, the numerical simulation matches closely the experimental values. The reattachment shock wave is reflected several times at the engine walls causing additional peaks in the pressure coefficient and the Stanton number. The impingement of the reflected reattachment shock at the lower engine wall is predicted too early by the RSM-Tr model. This might be caused by a too early laminar-to-turbulent transition of the boundary layer on the upper wall. Compared to the fully turbulent computation, where a fully turbulent boundary layer is assumed from the beginning of the upper wall, the shock impingement is shifted downstream and, thus, the prediction is improved. 

\subsection{DLR Intake}
\begin{figure}[htbp]
\begin{center}
 \includegraphics[height=0.3\textwidth]{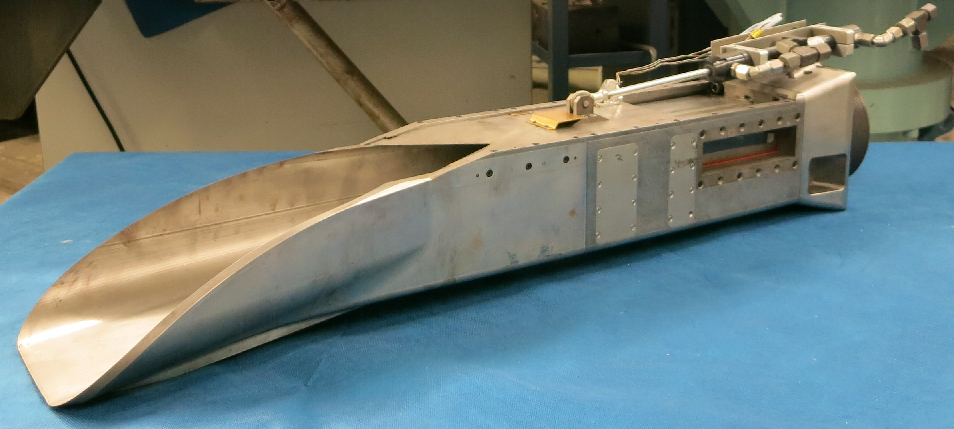}
\end{center}
 \caption{Photo of the DLR intake. Provided by A. Flock, DLR Cologne.}
\label{cad}
\end{figure}

This scramjet intake was designed and experimentally investigated at the German Aerospace Center (DLR) in Cologne~\cite{Hohn:12} and is referred to as DLR intake.
A photo of the intake presented within this section is shown in Figure \ref{cad}. The model is 750 mm long. The ramp angle is 8 degrees and the side wall angle 7 degrees each. The intersection of the ramp and side walls exhibits a smooth curvature. The sweep angle of the side wall is 45 degrees and reduces smoothly downstream. At  $x=650$~mm, the interface of the intake and the combustor is defined. Downstream of this location the walls are divergent by 1 degree. To reduce the spillage, a V-shaped lip is used. The front part of the cowl is movable to improve the starting behaviour and to adjust the intake to different Mach numbers. Thus, the lip position can be varied between $x_{lip}=300$~mm and $x_{lip}=450$~mm. At $x=550$~mm there is a small kink because the fixed part of the cowl starts. The numerical computations are performed for lip position $x_{lip}=330$~mm.

The configuration was tested at the Hypersonic Windtunnel H2K at the German Aerospace Center in Cologne. The test conditions in the experiments are listed in Table \ref{tbl:conditions}. These values are used as inflow conditions in the simulations. During the experiments, the surface pressure in the intake was measured by a total of 55 Kulite pressure probes \cite{Hohn:12}.
\begin{table}[htbp]
   \centering
        \begin{tabular}{|c|c|c|c|c|c|}\hline
$M_\infty$ [-]&$Re_\infty$~[$10^6$/m]&$T_0$~[K]&$T_\infty$[K]&$T_{\rm w}$~[K]&$I_\infty$~[\%]\\ \hline
7.0     				&  2.6                			&  700     		&     64.8   		  &    300 					&0.2  \\ \hline
   \end{tabular} \caption{Wind tunnel conditions for the DLR intake configuration.} \label{tbl:conditions}
\end{table}

For the numerical analysis, we perform multi-level computations using uniformly refined grids. The multi-level computations start on the refinement level $l=1$ until a density residual of $10^{-4}$ is reached. Then, this intermediate solution is used to initialize the next refinement level. This is repeated until a solution at the final level $l=4$ is achieved. The minimum wall distance of the final level $l=4$ grid is $1\times10^{-6}$~m which corresponds to a $y^+<1$. The final grid on level $l=4$ contains 13.5 million cells. Due to the computational effort no grid convergence study using a level $l=5$ grid (108 million cells) is performed. More information about the grid can be found in \cite{Frauholz:13a}.

\begin{figure*}[!ht]
\begin{center}
   \includegraphics[width=0.7\textwidth,clip]{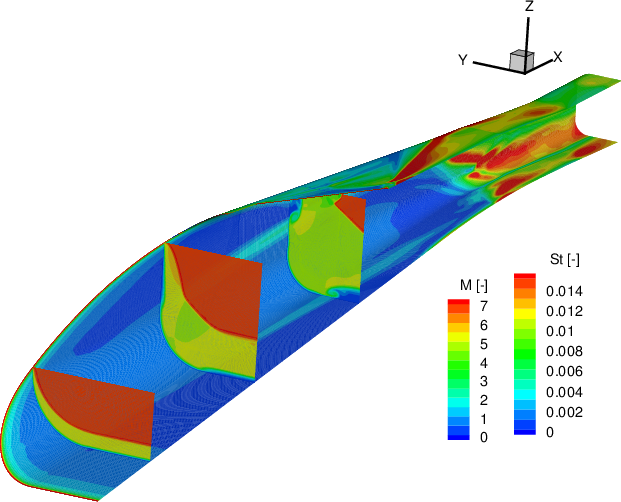}
   \caption{Mach number distribution at different cross sections of the half-model and Stanton number distribution at the intake walls for the DLR intake.}
   \label{3D_StM_base_symplane}	
\end{center}
\end{figure*}

To illustrate the three-dimensionality of the flow, Figure \ref{3D_StM_base_symplane} presents the normalized wall heat flux in terms of the Stanton number at the wall. The Mach number at different cross sections is also shown. The heat load at the exterior portion  of the intake is moderate, except for the leading edges of the ramp and the lower side wall. As the flow moves inside the intake, more shock waves are generated by deflection and impinge on the surface, creating several areas of intense heating.
The Mach number plots show the strong interaction of the leading edge shock wave and the side wall shock wave. Both shock waves are of approximately the same strength due to similar deflection angles. In the last cross section, a third shock wave, generated by the V-shaped cowl, appears and intensifies the interaction. Thus, the flow is highly three-dimensional.

\begin{figure*}[!ht]
\begin{center}
   \includegraphics[width=1\textwidth]{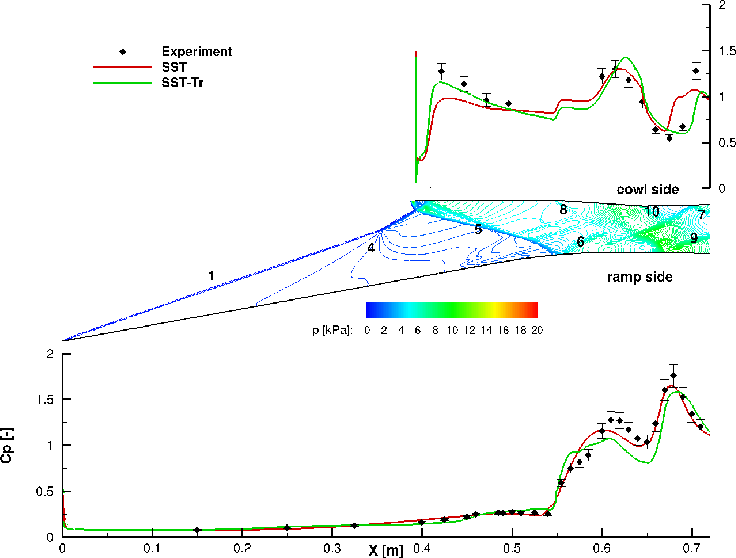}
   \caption{Pressure coefficient at the walls in the symmetry plane.}
   \label{cp_trans_symplane}	
\end{center}
\end{figure*}
\begin{figure*}[!ht]
\begin{center}
   \includegraphics[width=1\textwidth]{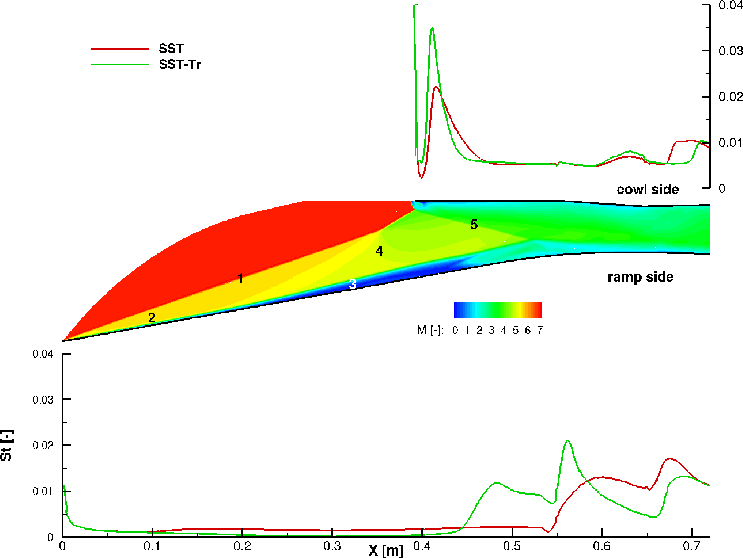}
   \caption{Stanton number at the walls in the symmetry plane.}
   \label{st_trans_symplane}	
\end{center}
\end{figure*}
To further analyse the flow, the distributions of the pressure coefficient and the Stanton number along the centerline are shown for both intake walls in Figure \ref{cp_trans_symplane} and \ref{st_trans_symplane}. There, we also show the pressure distribution and Mach number distribution in the symmetry plane, respectively. 
The incoming flow is compressed by the leading edge shock (1) which is visible in the pressure and Mach number lines. Due to the adverse pressure gradient that can be clearly seen in the pressure coefficient, the laminar boundary layer at the lower intake wall (2) thickens and separates (3) at $x\approx 0.25$~m. At $x\approx 0.45$~m, the flow reattaches producing a significant increase of the heat and pressure loads. Along the first ramp, the measured pressure data is closely matched by both numerical computations. However, due to the fully turbulent boundary layer predicted by the SST model, the heat loads are higher than for the SST-Tr model along the first ramp. In addition, the SST model does not predict the flow separation. Thus, the heat loads are not increased at $x\approx 0.45$~m (reattachment point for SST-Tr) for the SST model. 
Outside the boundary layer (4) compression lines caused by the side wall compression are visible. 
Due to the chosen off-design lip-position, the leading edge shock (1) hits the upper engine wall and interacts there with the lip shock (5) and the boundary layer. In this region, the advantages of the SST-Tr model can be clearly seen. Whereas the SST model is not able to predict the first pressure peak, the SST-Tr model follows closely the experimental data. Reason for this is the laminar onset of the cowl boundary layer. The peak heat transfer predicted by the SST model is only 60\% of the heat transfer predicted by the SST-Tr model. This shows, the importance of a correct prediction of the laminar-to-turbulent transition, especially for the wall heat loads.
The cowl shock wave (5) impinges at the lower intake wall causing the next increase of the pressure and the heat loads. The peak heating corresponding to shock impingement  is higher for the SST-Tr model, since the boundary layer is thinner in this region. The reflected shock wave (6) impinges at the upper intake wall and is there reflected (7).
At the kink between the movable and fixed part ($x=550$~mm) of the cowl, an oblique shock (8) is generated and reflected at the lower intake wall (9). After the interface to the combustor ($x=650$~mm), the intake divergences slightly and at the upper intake wall, the interaction of the shock reflection and the expansion (10) can be seen. 

\begin{figure*}
\begin{center}
\includegraphics[width = 0.7\textwidth,  clip]{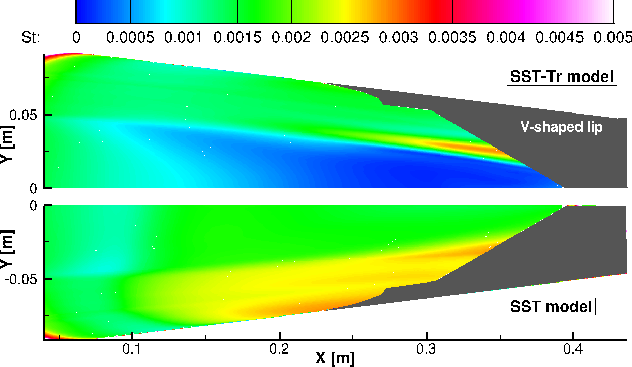} \\
    \includegraphics[width = 0.7\textwidth,  clip]{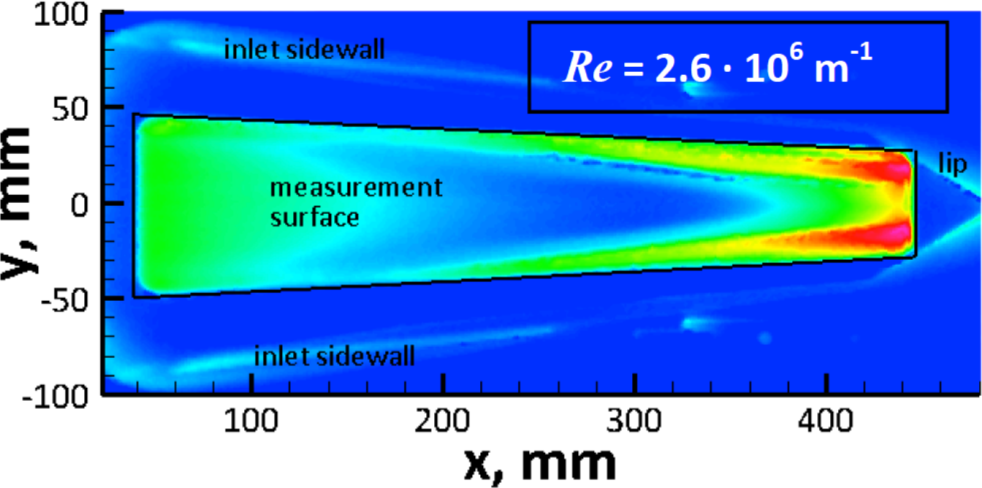} 
  \caption{Stanton number distribution at the compression ramp for the computations (top) and experiments (bottom). Experimental data reproduced from~\cite{Hohn:12}.}\label{dlr-st-ramp-scram}
\end{center}
\end{figure*}
Figure \ref{dlr-st-ramp-scram} shows the Stanton number distribution at the external compression ramp for the computations (top) and the experiment (bottom). The lip position during the experiment differs from the lip position of the numerical computations, but this does not influence the external ramp flow. The fully turbulent SST model strongly overpredicts the measured heat loads and is not able to precisely predict the heat transfer. In contrast, the SST-Tr model accurately captures the transition process resulting in a good agreement with the measured heat loads at the external ramp. Thus, the using the transition model strongly improves the accuracy of the numerical results compared to the experimental data.
 
\subsection{ITAM Intake}
\begin{figure}[htbp]
\begin{center}
 \includegraphics[height=0.3\textwidth]{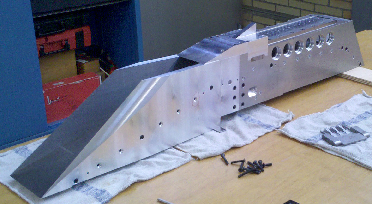}
\end{center}
 \caption{Photo of the ITAM intake.}
\label{itam-cad}
\end{figure}

A photo of the intake presented within this section is shown in Figure \ref{itam-cad}. 
The intake was tested at the Hypersonic Windtunnel IT302 of the Institute of Theoretical and Applied Mechanics (ITAM) in Novosibirsk, Russia. The test conditions in the experiments are listed in Table \ref{tbl:conditions}. These values are used as inflow conditions in the simulations. Up to now, the experimental data are not published.
\begin{table}[htbp]
   \centering
        \begin{tabular}{|c|c|c|c|c|c|}\hline
$M_\infty$&$Re_\infty$~[$10^6$/m]&$T_0$~[K]&$T_\infty$[K]&$T_{\rm w}$~[K]&$I_\infty$~[\%]\\ \hline
8.0     &  2.66                &  3280     &     237.7   &    300 &0.5 \\\hline
   \end{tabular} \caption{Wind tunnel conditions for the ITAM intake configuration.} \label{tbl:conditions}
\end{table}
This intake contains a single exterior compression ramp with a deflection angle of 15.5 degrees and a straight interior part. The overall length is 580~mm. The side walls have a sweep angle of 35 degrees and a compression angle of 3.5 degrees. The cowl lip starts at $x=400$~mm. The height of the interior section is 34~mm. 

\begin{figure}[htbp]
\begin{center}
 \includegraphics[width=0.49\textwidth,clip]{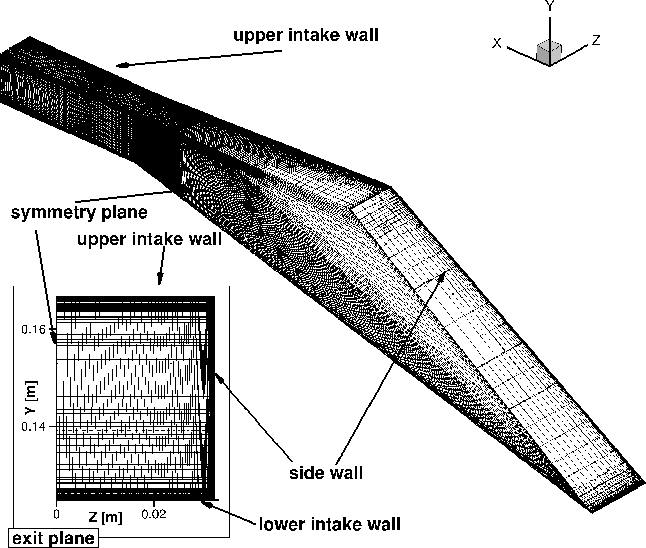}
  \includegraphics[width=0.49\textwidth ,clip]{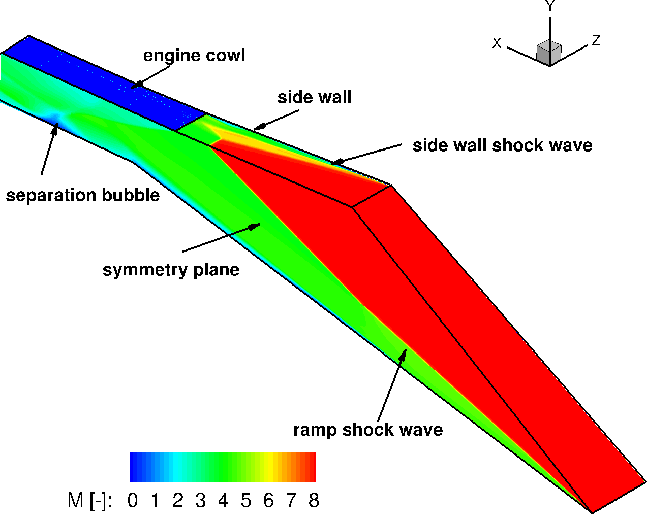}
\end{center}
 \caption{Left: Final adaptive grid at level $l=4$ of the ITAM intake. Right: Mach number distribution of the ITAM intake for the RSM-Tr model. }
\label{itam-grid}
\end{figure}

For the numerical analysis, we perform adaptive computations using 4 refinement levels. 
On refinement level $l=0$, the grid has 24 cells in the flow direction, 9 cells in the cross-flow direction and 6 cells in spanwise-direction. The final adaptive level $l=4$ grid has a minimum wall distance of $1\times10^{-6}$~m to resolve correctly the strong gradients within the boundary layer and it contains 4.4 million cells. Due to the computational effort, no grid convergence study is performed. Only the half-model is computed using a symmetry boundary condition. Figure \ref{itam-grid} (left) shows the final level $l=4$ grid. Information about previous numerical studies can be found in \cite{Rein:08,Reinartz:12}.

The computation of the RSM-Tr model is used to visualize the general flow features in Figure \ref{itam-grid} (right). 
The incoming flow is compressed and decelerated by the strong ramp shock wave and by the weaker side wall shock wave. When entering the interior part the flow is turned into the engine and the flow expands. The lip shock of the engine cowl is deflected by the expansion fan  at the lower intake wall. The impingement of the lip shock at the lower engine wall causes a separation bubble. 

\begin{figure}[htbp]
\begin{center}
 \includegraphics[width=1\textwidth,clip]{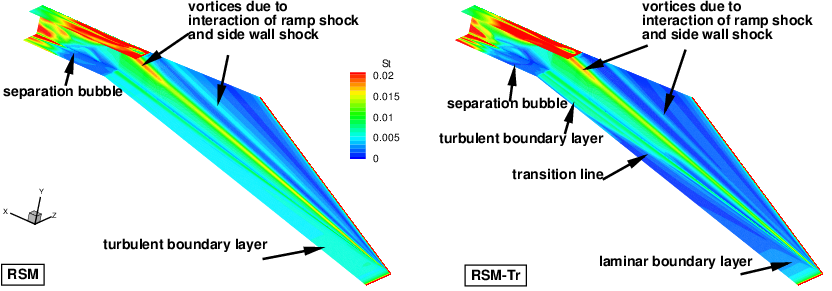}
\end{center}
 \caption{Stanton number distribution of the ITAM intake. }
\label{itam-3d-st}
\end{figure}

To illustrate the three-dimensional effects in the flow field, Figure \ref{itam-3d-st} shows the footprint of the flow structures on the ramp and the side wall in terms of Stanton number for the RSM model and the RSM-Tr model. Compared to the fully turbulent computation using the RSM model, the heat loads at the ramp are smaller for the RSM-Tr model since the boundary layer is laminar. The laminar boundary layer thickens and transitions to turbulent at the ramp where the Stanton number reaches the turbulent level. The flow transitions first close to the side wall due to interactions of the side wall shock wave with the ramp shock wave. The footprints of the vortices generated by this interaction can be seen in the lines with increased Stanton number at the ramp and the side wall. For both models the lip shock wave and the 3D vortices lead to high heat loads at the engine cowl. The separation bubble produced by the lip shock impinging at the boundary layer reduces the heat loads in the separated flow region at the lower engine wall.

\begin{figure}[htbp]
\begin{center}
 \includegraphics[width=1\textwidth]{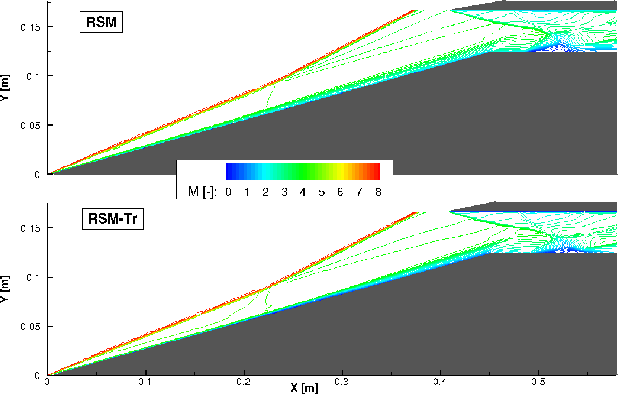}
\end{center}
 \caption{Mach number lines in the symmetry plane of the ITAM intake.}
\label{itam-2d-m}
\end{figure}

To further investigate the flow separation in the interior and the interaction of the lip shock wave with the expansion fan, Figure \ref{itam-2d-m} shows the Mach number lines in the symmetry plane for the RSM model and the RSM-Tr model. 
At the first part of the ramp, the laminar boundary layer for the RSM-Tr model is thinner than the fully turbulent boundary layer computed by the RSM model. The laminar boundary layer thickens strongly after $x\approx0.25$~m. Especially the subsonic part of the boundary layer rapidly grows. A closer look at the velocity and pressure profiles (not shown here) proved that the boundary layer is highly distorted and nearly separated. Thus, at the end of the ramp the boundary layer is nearly as thick as the fully turbulent boundary layer of the RSM model. After the expansion corner, the boundary layer computed is thinner for the RSM-Tr model resulting in a smaller separation bubble. 

\begin{figure}[htbp]
\begin{center}
 \includegraphics[width=0.7\textwidth,clip]{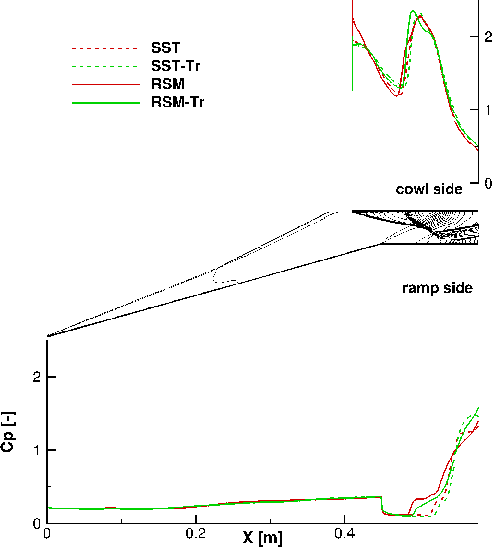} 
\end{center}
 \caption{Pressure coefficient distribution in the symmetry plane of the ITAM intake. The pressure isolines for the RSM-Tr model are shown as well. }
\label{itam-cp}
\end{figure}
\begin{figure}[htbp]
\begin{center}
 \includegraphics[width=0.7\textwidth,clip]{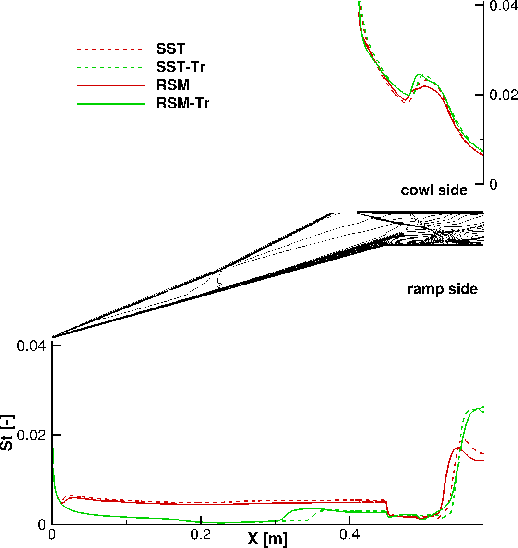}
\end{center}
 \caption{Stanton number distribution in the symmetry plane of the ITAM intake. The Mach number isolines for the RSM-Tr model are shown as well.}
\label{itam-st}
\end{figure}

Figure \ref{itam-cp} and \ref{itam-st} show the pressure coefficient and the Stanton number distribution at the wall along the centerline of the engine for all four models. The laminar boundary layer computed by the SST-Tr and RSM-Tr model result in lower heat loads along the external ramp. The flow expansion can be seen in the drop of pressure and heat loads for all computations. The lip shock increases the pressure and heat loads at the upper wall. Due to the impinging lip shock wave at the lower engine wall the flow separates. 
Table \ref{tbl:sep-size-itam} lists the location and size of the separation bubble occurring at the lower engine wall for the different computations.
The largest separation (0.0378 m) is predicted by the RSM model, whereas the SST-Tr model predicts the smallest separation length (0.0167 m). 
Compared to the fully turbulent computations, the separation point is moved upstream and the separation length is significantly reduced for the computations using the transition model.
The reattachment shock produces peaks in the pressure and heat loads. The second peak of the pressure coefficient and the Stanton number at the cowl side is a result of the 3D vortices generated by the interaction of the ramp shock and the side wall shock.
Besides the strong impact of the transition model on the prediction of the heat loads, this test case shows the importance of an accurate transition prediction regrading the location and size of the flow separation. The flow separation influences the captured mass flow and, hence, the engine performance.

\begin{table}[htbp]
   \centering
        \begin{tabular}{|c|c|c|c|c|}\hline
&separation point  &reattachment point &separation length\\ \hline
SST     &     0.5179  m     &    0.5412 m  & 0.0233 m \\ \hline
SST-Tr     &     0.5245  m     &    0.5412 m & 0.0167 m   \\ \hline
RSM     &     0.4884  m     &    0.5262 m  &  0.0378 m\\ \hline
RSM-Tr     &     0.4959  m     &    0.5331 m &  0.0372 m \\ \hline
   \end{tabular} \caption{Location and size of the separation bubble at the lower engine wall for the ITAM intake.} \label{tbl:sep-size-itam}
\end{table}

\section{Conclusions}
Within this paper, we successfully coupled the $\gamma$-Re$_{\theta_t}$ transition model of Langtry/Menter to Eisfeld's RSM model. First tests showed the necessity of calibrating the empirical correlation for hypersonic flow regime. After calibrating using a hypersonic flat plate, we successfully validate the RSM-Tr model with a hypersonic double ramp test case. Within previous work, the SST model was coupled to the $\gamma$-Re$_{\theta_t}$ transition model. For completeness, the validation of the SST-Tr model is shown as well. Then, three-dimensional computations are performed using the RSM-Tr model and the SST-Tr model for three different scramjet intakes. In contrast to fully turbulent computations, the use of the transition model results in a better agreement with the available  experimental data, since the boundary starts laminar and transitions with respect to the flow physics. 
Compared to the fully turbulent computations, the computations using the transition model predict lower heat loads in the region of the laminar boundary layer. 
For the SWL intake the separation bubble at the external ramps is only predicted when using the transition model. The shock waves in the interior part of the engine are shifted upstream when usinf the transition model. Although the transition process is predicted slightly too early the agreement to the measured pressure and heat loads is improved.
The DLR intake shows the advantages of using the transition model as well. Especially at the upper engine wall only the computation using the transition model is able to precisely predict the measured pressure loads. In addition, the heat loads at the external ramp are only accurately captured by the computation using the transition model.
For the ITAM intake the size and the location of the separation bubble in the interior part is significantly reduced by the transition model showing the importance of modelling the transition process. 

\section*{Acknowledgments}
This work was supported by the German Research Foundation (DFG) within the framework of the GRK 1095 ``Aero-Thermodynamic Design of a Scramjet Propulsion System for Future Space Transportation Systems'' and the GSC 111 Aachen Institute for Advanced Study in Computational Engineering Science (AICES). Computing resources were provided by the RWTH Aachen University Center for Computing and Communication and the Forschungszentrum J\"ulich. Thanks to the Shock Wave Laboratory, RWTH Aachen University and the German Aerospace Center in Cologne for providing the experimental data.

\section*{References}
\bibliography{flower}
\bibliographystyle{aiaa}

\end{document}